\documentclass[aps,10pt,reprint,amsmath,amssymb,pra,superscriptaddress]{revtex4-2}

\usepackage[utf8]{inputenc}
\usepackage{xcolor}
\usepackage{graphicx}
\usepackage{subfigure}
\usepackage{bookmark}
\usepackage[normalem]{ulem}
\usepackage{comment}
\usepackage{dsfont} 
\usepackage[ruled,linesnumbered]{algorithm2e}
\usepackage[skip=0pt plus0pt, indent=10pt]{parskip}
\usepackage{physics}
\usepackage{xfrac}
\usepackage{here}
\usepackage{hyperref}
\usepackage{siunitx}
\usepackage{natbib}
\usepackage{bm}

\usepackage{hyperref}
\hypersetup{
    colorlinks=true,
    linkcolor=blue,
    filecolor=blue,      
    urlcolor=blue,
    citecolor=blue
}

\newcommand{\comma}{~,}
\newcommand{\fullstop}{~.}
\DeclareRobustCommand{\Chi}{{\mathpalette\irchi\relax}}
\newcommand{\irchi}[2]{\raisebox{\depth}{$#1\chi$}}
        
\begin{document}

\title{Depth Determination of Individual Shallow NV-Centers via Spin-Lock NMR}

\author{Aaron Daniel}
\thanks{These authors contributed equally.}
\affiliation{Department of Physics, University of Basel, Klingelbergstrasse 82, CH-4056 Basel, Switzerland}

\author{Beat Bürgler}
\thanks{These authors contributed equally.}
\affiliation{Department of Physics, University of Basel, Klingelbergstrasse 82, CH-4056 Basel, Switzerland}

\author{Patrick Maletinsky}
%\email{patrick.maletinsky@unibas.ch}
\affiliation{Department of Physics, University of Basel, Klingelbergstrasse 82, CH-4056 Basel, Switzerland}

\author{Patrick Potts}
\thanks{Now at ABB Corporate Research, Segelhofstrasse 1K, 5405 Baden-Dättwil, Switzerland}
\affiliation{Department of Physics, University of Basel, Klingelbergstrasse 82, CH-4056 Basel, Switzerland}

\date{\today}

\begin{abstract}
Quantitative quantum sensing with shallow electron spins, such as those hosted by nitrogen-vacancy (NV) centers in diamond, requires accurate knowledge of the spin's depth below the host material's surface.
A widely used approach infers this depth from the $^{1}$H nuclear magnetic resonance~(NMR) signal of immersion oil on the diamond surface that can be detected using dynamical decoupling sequences such as XY8.
However, finite-width pulses make XY8 sensitive to subharmonic responses, including unwanted contributions from nearby \(^{13}\mathrm{C}\) spins, and its instrument-limited spectral resolution provides only sparse sampling of the narrow $^{1}$H NMR lineshape.
Here, we introduce Spin-Lock NMR as an alternative approach to single-NV depth determination.
By tuning the Spin-Lock Rabi frequency to the $^{1}$H Larmor frequency, the NV probes the $^{1}$H NMR signal through the Hartmann--Hahn resonance without the harmonic ambiguities of pulsed decoupling sequences and with substantially higher instrument-limited spectral resolution.
We derive a quantitative Spin-Lock NMR fit function from a Markovian master equation that directly relates the measured spectrum to the NV depth.
Our approach yields NV depth estimates in excellent agreement with the established XY8-based protocol across multiple NV centers and establishes Spin-Lock NMR as a robust alternative for quantitative single-NV depth determination.
To demonstrate its applicability, we employ our method to investigate the $^{1}$H nuclear spin signal that is regularly reported to be present on diamond, even in the absence of immersion oil. 
\end{abstract}

\maketitle

\section*{Introduction}

% ---------------------------------------------------------
% INTRODUCTION TO NV QUANTUM SENSING
% ---------------------------------------------------------
The nitrogen-vacancy (NV) center is a point defect in diamond~\cite{Doherty2013a} that has emerged as a powerful spin-based quantum sensor, with demonstrated and impactful applications in, for instance, the life sciences~\cite{Barry2016a}, geological sciences~\cite{Glenn2017a}, materials science~\cite{Casola2018a} and navigation~\cite{Wang2024a}.
The key properties that render the NV center such an exceptional quantum sensing platform are the long coherence time of its intrinsic electron spin even at room temperature~\cite{Balasubramanian2009a,BarGill2013a,Kennedy2003a}, polarization and readout of the spin state by purely optical means~\cite{Tetienne2012a,Harrison2004a}, and the highly robust and chemically inert diamond host material that is not only compatible with in-vivo sensing~\cite{Fujiwara2020a}, but that can also be nanofabricated into experimentally advantageous nanostructures such as scanning probes for scanning NV magnetometry~(SNVM)~\cite{Maletinsky2012a}.

% ---------------------------------------------------------
% IMPORTANCE OF THE NV DEPTHS IN SENSING APPLICATIONS
% ---------------------------------------------------------
For most NV sensing applications, shallow NV centers located only a few nanometers below the diamond surface are especially useful because they can be brought into close proximity to the sensing target, which enhances both signal strength and spatial resolution.
Importantly, the quantitative performance of these applications often depends critically on accurate knowledge of the NV depth in the diamond crystal.
In SNVM applications, for example, the quantitative reconstruction of the sample magnetization from the measured magnetic stray fields is explicitly dependent on the NV depth~\cite{Thiel2019a,Dubois2022a,Xu2025a}.

% ---------------------------------------------------------
% EXPLANATION WHY SRIM IS NOT GOOD ENOUGH
% ---------------------------------------------------------
For NV centers formed by ion implantation, a common tool for depth estimation is the stopping range of ions in matter (SRIM) Monte-Carlo simulation~\cite{Ziegler2012a}.
However, SRIM provides only a statistical prediction and not a precise measurement of an individual NV center's depth.
Moreover, it neglects certain effects such as crystal channeling of ions~\cite{Lehtinen2016a}, causing it to underestimate the implantation depth by up to a factor of two~\cite{Toyli2010a}.
This discrepancy is particularly pronounced for NV centers in scanning probe tips, where SRIM consistently predicts depths even less than half of the NV-sample distances inferred from SNVM magnetization reconstruction~\cite{Appel2019a,PelletMary2025a,Xu2025a}.
Consequently, SRIM does not satisfy the need for a quantitatively accurate depth determination technique.

\begin{figure}[t]
\includegraphics{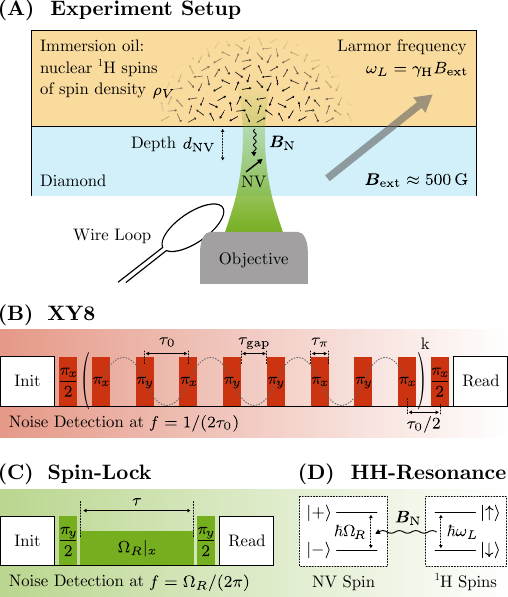}
\caption{\label{Fig1} 
\textbf{Experimental arrangement for depth determination of single NV centers}.
\textbf{(A)}~The diamond surface is coated with immersion oil which contains $^{1}$H nuclear spins at a density~$\rho_V$.
In an external magnetic field $\bm{B}_{\rm ext}$ these nuclear spins produce magnetic fluctuations~$\bm{B}_{\rm N}$ that peak at the Larmor frequency $\omega_L$. 
By applying a suitable NMR control sequence with a gold wire loop, these fluctuations can be detected by a shallow NV spin.
Since the $^{1}$H spin noise amplitude depends on the NV depth~$d_{\rm NV}$, the latter can be determined by fitting the NMR signal with a quantitative theoretical model.
\textbf{(B)}~The established sequence for such $^{1}$H NMR is XY8, a dynamical decoupling sequence comprised of a series of phase-cycled $\pi$-pulses. 
It enables magnetic noise detection at a frequency $f=1/(2\tau_0)$, where $\tau_0$ is the center-to-center spacing between consecutive $\pi$-pulses.
\textbf{(C)}~Spin-Lock is an alternative to XY8 for NMR, where the NV spin is driven with a phase-matched pulse of Rabi frequency~$\Omega_R$. 
\textbf{(D)}~Such Spin-Lock NMR is based on the Hartmann-Hahn~(HH) resonance condition, where the Spin-Locked system -- characterized by its driven eigenstates $\ket{\pm}$ with transition frequency $\Omega_R$ -- is tuned into resonance with the Larmor frequency, $\Omega_R=\omega_L$, to enable detection of the nuclear magnetic noise field $\bm{B}_{\rm N}$, with a detection-frequency window centered at $f=\Omega_R/(2\pi)$.
}
\end{figure}

% ---------------------------------------------------------
% INTRODUCTION TO PHAM'S XY8 DEPTH DETERMINATION APPROACH
% ---------------------------------------------------------
In 2016, Pham \textit{et al.}~\cite{Pham2016a} introduced a promising method for determining the depth of individual NV centers.
Their approach relies on measuring the nuclear magnetic resonance~(NMR) of $^{1}$H nuclear spins in immersion oil deposited on the diamond surface (Fig.~\ref{Fig1}\,(A)).
The NMR signal is detected using an XY8 dynamical decoupling sequence~\cite{Gullion1990a}, which consists of a series of phase-cycled microwave $\pi$-pulses separated by a delay~$\tau_0$~(Fig.~\ref{Fig1}\,(B)).
Together, these pulses effectively act on the NV spin like a narrow band-pass noise filter~\cite{Biercuk2011a} centered at frequency $f=1/(2\tau_0)$.
By sweeping $f$ across the $^{1}$H Larmor frequency~$\omega_L$, an NMR spectrum is obtained~\cite{Bucher2019a,Allert2022a,Liu2022a}, and by carefully modeling the interaction of the NV spin with the magnetic signal~$\bm{B}_{\rm N}$ produced by the $^{1}$H spins (assuming a known layer thickness and nuclear spin density~$\rho_V$), one can fit said spectrum and thereby infer the depth of the deployed NV defect.

% ---------------------------------------------------------
% FIGURE 1
% ---------------------------------------------------------

% ---------------------------------------------------------
% WEAKNESSES OF XY8 FOR NMR DEPTH DETERMINATION
% ---------------------------------------------------------
However, as we will show below, the XY8 sequence comes with two critical weaknesses that complicate this depth determination scheme:
First, realistic $\pi$-pulses of non-zero duration~$\tau_\pi$ make the sequence sensitive to subharmonics of the target frequency as well as other spurious magnetic signals~\cite{Loretz2015a}.
For $^{1}$H NMR specifically this means that XY8 is not only collecting the desired $^{1}$H signal, but also the fourth harmonic of $^{13}$C spins in the diamond lattice, ultimately leading to an underestimation of the NV depth when using the model of Pham \textit{et al.}
Second, XY8 NMR suffers from a critically low instrument-limited frequency resolution.
Depending on the pulse timing resolution of the deployed hardware and the applied magnetic bias field~$B_{\rm ext}$, typically achievable NMR resolutions lie in the range of $<$\SI{1}{kHz} to $\sim$\SI{10}{kHz}.
Given typical $^{1}$H NMR linewidths of 10 to \SI{40}{kHz}~\cite{Pham2016a}, this might lead to undersampling of the NMR line and thereby reduce the precision of the resulting depth fit.

% ---------------------------------------------------------
% AVENUES TO OVERCOME XY8 WEAKNESSES
% ---------------------------------------------------------
There are two principal strategies to overcome these problems with XY8 NMR: tailoring the experimental conditions or modifying the XY8 sequence itself.
For the former, one can for instance work with isotopically purified diamond to eliminate the $^{13}$C signal all together, however, this restricts this type of depth determination to specific diamond materials.
Alternatively, one can operate at lower magnetic bias fields~$B_{\rm ext}$, which improves the attainable spectral resolution and reduces the effect of the $^{13}$C subharmonics by minimizing the ratio $\tau_\pi/\tau_0$.
The drawback of this approach, however, is that it forfeits the advantages of operating near the NV excited state level anti-crossing (ESLAC) at $B_{\rm ext}\approx\SI{500}{G}$, where hyperpolarization of the NV center's nuclear nitrogen spin~\cite{Jacques2009a} suppresses interference from off-resonantly driven hyperfine transitions, and simultaneously enhances the optical contrast and thus the signal-to-noise ratio (SNR) in NMR measurements.
Possible modifications of the XY8 sequence that address its weaknesses are XY8 correlation spectroscopy, whose spectral resolution depends only on the NV spin lifetime~\cite{Boss2017a,Jiang2023a}, adaptive XY sequences with non-equally spaced pulses that eliminate low-order harmonics~\cite{Casanova2015a}, controlled randomization of the $\pi$-pulse phase to mitigate spurious signals~\cite{Wang2019a,Wang2020a}, or so-called cyclic geodesic driving which generates repeated adiabatic evolutions of the NV spin to suppress both harmonics and other spurious signals~\cite{Zeng2024a}.
However, these modifications come at the cost of increased pulse complexity, longer sequence durations, and consequently a reduced SNR.

% ---------------------------------------------------------
% INTRODUCE OUR SPIN-LOCK APPROACH TO DEPTH DETERMINATION
% ---------------------------------------------------------
Here, we present a fundamentally different approach to measuring the $^{1}$H NMR of immersion oil on the diamond surface for the quantitative depth determination of single NV centers.
Specifically, we propose to use the Spin-Lock sequence instead of XY8, and will show (cf. Sec.~\ref{sec:spin_lock_advantages}) that, without any drawbacks, Spin-Lock NMR overcomes the weaknesses of XY8 NMR in that spurious harmonics are entirely absent and a high spectral resolution better than \SI{0.5}{kHz} can be achieved at all magnetic bias fields.
While Spin-Lock NMR has been discussed previously~\cite{Loretz2013a,Yan2013a,Rosskopf2014a,Kost2015a,Hermann2024a}, the key novelty of our work (Sec.~\ref{sec:depth_spinlock_nmr}) is a detailed and rigorous theoretical framework to describe the interaction of the Spin-Locked NV spin with an ensemble of $^{1}$H spins on the diamond surface, that enables quantitative depth determination via Spin-Lock NMR.
The depth values that we obtain this way are consistent with the results of the XY8 protocol across multiple NVs.
We therefore conclude that our Spin-Lock approach represents a promising alternative to the established XY8-based depth determination scheme, that brings clear advantages for NV depth determination in a range of application regimes and relevant experimental settings.
Finally, we apply our method (Sec.~\ref{sec:adsorbate_layer}) to characterize a ubiquitous $^{1}$H-containing adsorbate layer present on clean diamond surfaces that has previously been observed~\cite{Abendroth2022a,Xu2025a,Xu2026a,DeVience2015a,Loretz2014a,Rugar2015a,Zheng2026a,Bruckmaier2023a}.
Our analysis suggests that water alone cannot account for this signal, and we estimate that its presence does not significantly affect immersion-oil-based NV depth determination approaches.

% ---------------------------------------------------------
% FIGURE 2
% ---------------------------------------------------------
\begin{figure*}[t]
\includegraphics{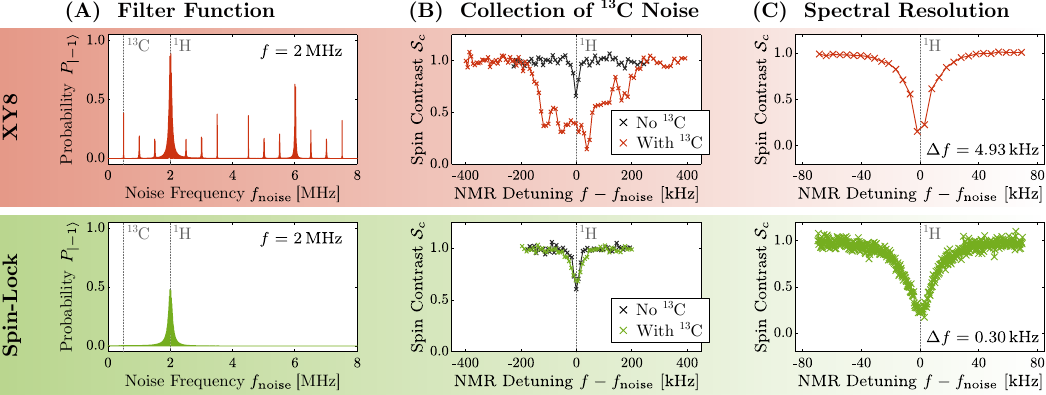}
\caption{\label{Fig2} 
\textbf{Comparison of XY8 and Spin-Lock for $^{1}$H NMR}.
\textbf{(A)}~Numerically computed filter functions of these two sequences, describing the probability $P_{\ket{-1}}$ of the NV spin to be in state $\ket{-1}$ at the end of the sequence given that it was initially prepared in~$\ket{0}$, and plotted as a function of the applied noise frequency $f_{\rm noise}=\omega_L/(2\pi)$.
Each sequence is constructed such that its detection frequency $f$ is \SI{2}{MHz}~(=~resonant with $^{1}$H spins at $B_{\rm ext}\approx\SI{500}{G}$).
Our simulation shows that XY8, simulated for realistic pulses with $\tau_\pi/\tau_0=\SI{20}{\percent}$, collects not only the $^{1}$H noise at \SI{2}{MHz} but also noise at various harmonics of $f$.
In particular, there is a subharmonic on resonance with $^{13}$C noise (\SI{0.5}{MHz}).
Spin-Lock, on the other hand, shows no harmonic side peaks at all and does therefore only collect the desired $^{1}$H signal.
\textbf{(B)}~NMR data taken with XY8 and Spin-Lock on two different, single NV defects. One defect is located in an isotopically purified diamond with negligibly little $^{13}$C, the other contains a natural abundance of $^{13}$C.
Quite clearly, Spin-Locking is insensitive to $^{13}$C, while XY8 is detecting these unwanted $^{13}$C-related fluctuations. 
\textbf{(C)}~Immersion oil $^{1}$H NMR data taken with XY8 at a field of $B_{\rm ext}=\SI{519.6}{G}$ and with the highest possible spectral resolution available to us at this field (our hardware pulse timing resolution is $\Delta t = \SI{0.5}{ns}$), yielding $\Delta f = \SI{4.93}{kHz}$, which matches the prediction of Eq.~\eqref{eq:spectralresolution}.
With Spin-Lock NMR, one can achieve a much higher resolution limited only by the accuracy with which the microwave amplitude can be controlled.
Here, we present Spin-Lock NMR data taken on the same NV and at the same magnetic field, with a spectral resolution of $\Delta f=\SI{0.30}{kHz}$, exceeding the XY8 resolution limit by a factor of more than ten.
}
\end{figure*}

% ---------------------------------------------------------
% Section I
% ---------------------------------------------------------
\section{Advantages of Spin-Lock NMR over XY8-based NMR}
\label{sec:spin_lock_advantages}

% ---------------------------------------------------------
% DESCRIPTION OF SPIN-LOCK NMR
% ---------------------------------------------------------
The Spin-Lock sequence (see Fig.~\ref{Fig1}\,(C)) begins by optically initializing the NV spin into its~$\ket{0}$ spin state.
Here, $\ket{0}$ and $\ket{\pm1}$ denote the NV spin eigenstates of the $\hat{S}_z$ spin operator, where $z$ points along the NV symmetry axis.
A subsequent $\pi_y/2$ pulse prepares the equal superposition state $\ket{+}=(\ket{0}+\ket{-1})/\sqrt{2}$, after which a second microwave pulse, resonant with the NV $\ket{0}\leftrightarrow\ket{-1}$ transition and with Rabi frequency $\Omega_R$, is applied to the NV.
Importantly, the microwave phase of this pulse is chosen to match the quantum phase of~$\ket{+}$, i.e. it is phase-shifted by $\pi/2$ from the first pulse, such that in the rotating frame Bloch sphere picture, both the driving field and the NV spin state $\ket{+}$ point along the positive $x$-axis.
Consequently, during this so-called Spin-Locking pulse, $\ket{+}$ is an eigenstate of the driven system, while the orthogonal state $\ket{-}=(\ket{0}-\ket{-1})/\sqrt{2}$ forms the second eigenstate.
These dressed states $\ket{\pm}$ are separated by an energy~$\hbar\Omega_R$, which renders the Spin-Locked system primarily sensitive to magnetic fluctuations at frequency~$\Omega_R$, while magnetic noise at all other frequencies does not couple efficiently to the spin.
Consequently, the Spin-Lock sequence acts as a narrow band-pass noise filter centered at $f=\Omega_R/(2\pi)$. 
It therefore provides an alternative approach to frequency-selective noise spectroscopy, and in particular to NV NMR experiments where $\Omega_R$ is tuned to be resonant with the Larmor frequency $\omega_L$ of a nearby nuclear spin species.
This resonance condition, illustrated in Fig.~\ref{Fig1}\,(D), is the key mechanism of Spin-Lock NMR and is commonly referred to as the Hartmann--Hahn~(HH) resonance~\cite{HartmannHahn1962a}.

% ---------------------------------------------------------
% EXPLAIN OUR CHOICE OF MAGNETIC FIELD (500G)
% ---------------------------------------------------------
%In this work specifically, we probe $^{1}$H nuclear spins in an external magnetic bias field near the excited state level anti-crossing (ESLAC), $B_{\rm ext}\approx \SI{500}{G}$, where $^{1}$H Larmor frequency is roughly $\omega_L\approx2\pi\times\SI{2}{MHz}$. We make this choice of magnetic field for two reasons: First, due to the hyperpolarization of the NV center's nitrogen nuclear spin near the ESLAC~\cite{Jacques2009a}, interference from any off-resonantly driven hyperfine transitions can be avoided. Second, this isolated hyperpolarized transition has an increased optical contrast compared to an unpolarized line, leading to an enhanced signal-to-noise ratio in NMR experiments.

% ---------------------------------------------------------
% C13 HARMONICS: FILTER FUNCTIONS
% ---------------------------------------------------------
In contrast to XY8, Spin-Lock is not a periodic multi-pulse sequence and it is therefore inherently insensitive to harmonic frequencies.
To demonstrate this, we calculate the probability~$P_{\ket{-1}}$ of finding the NV spin in state~$\ket{-1}$ at the end of the Spin-Lock sequence, assuming the NV was initially prepared in~$\ket{0}$ and that it is subject to a nearby nuclear spin with Larmor frequency~$\omega_L$.
The calculation follows a slightly modified version of the model introduced in Ref.~\cite{Loretz2015a} and is presented in App.~\ref{app:filterfunc}.
To demonstrate the sensitivity to different noise frequencies, we evaluate $P_{\ket{-1}}$ as a function of the nuclear noise frequency~$f_{\rm noise}=\omega_L/(2\pi)$ and for a fixed detection frequency of $f=\SI{2}{MHz}$, corresponding to the $^{1}$H Larmor frequency at the magnetic bias field of $B_{\rm ext}\approx\SI{500}{G}$ used throughout this work.
The result of this (Fig.~\ref{Fig2}(A), bottom) can be interpreted as a noise filter function~\cite{Biercuk2011a}.
In addition, we calculate the same filter function for XY8, where we explicitly include realistic $\pi$-pulses with a non-zero duration of $\tau_\pi=\SI{50}{ns}$, such that $\tau_\pi/\tau_0=\SI{20}{\percent}$~(Fig.~\ref{Fig2}(A), top).
As previously demonstrated~\cite{Loretz2015a}, the XY8 filter function exhibits not only the main resonance at \SI{2}{MHz}, but also multiple harmonic responses, including a subharmonic at \SI{0.5}{MHz}.
For the experimentally relevant case in which the main resonance probes the $^{1}$H Larmor frequency~(\SI{2}{MHz}), this specific subharmonic coincides with the $^{13}$C Larmor frequency~(\SI{0.503}{MHz}), reconfirming that XY8-based $^{1}$H NMR with non-zero $\pi$-pulse durations is intrinsically sensitive to $^{13}$C spins.
In contrast, the Spin-Lock filter function contains only a single resonance at the target $^{1}$H frequency of \SI{2}{MHz} and no harmonic peaks at all, demonstrating complete immunity to any $^{13}$C-induced magnetic fluctuations.

% ---------------------------------------------------------
% C13 HARMONICS: EXPERIMENTAL DATA
% ---------------------------------------------------------
To provide experimental evidence for this immunity to $^{13}$C, we measure the NMR signal of $^{1}$H spins in an $^{1}$H-containing adsorbate surface layer on the diamond surface~\cite{Abendroth2022a,Xu2025a,Xu2026a,DeVience2015a,Loretz2014a,Rugar2015a,Zheng2026a,Bruckmaier2023a}.
This signal is measured with both XY8 and Spin-Lock, each on an NV in an isotopically purified diamond with negligibly little $^{13}$C ($<\SI{0.001}{\percent}$) and on an NV in a diamond crystal with a natural abundance of $^{13}$C ($\approx\SI{1.1}{\percent}$).
These NVs were created by nitrogen ion implantation (at an energy of 6 and \SI{12}{keV} respectively) and subsequent annealing (see App.~\ref{app:diamondsamples} for more details).
Here and elsewhere in this work, we plot our NMR data as a function of the NMR detuning $\Delta=2\pi\times(f-f_{\rm noise})$; and we normalize the data as described in App.~\ref{app:data_normalization} to obtain units of spin contrast~$\mathcal{S}_c\in[0,1]$, with 
\begin{align}
    \label{eq:spin_contrast_def}
    \mathcal{S}_c = P_{\ket{+}} - P_{\ket{-}}\comma  
\end{align}
where $P_{\ket{\pm}}$ is the probability for the system to be in $\ket{\pm}$ at the end of the Spin-Locking pulse or the XY8 $\pi$-pulse train respectively.
This normalization to $\mathcal{S}_c$ effectively removes any influence from noise frequencies outside the investigated spectral window.
The resulting NMR data, as shown in Fig.~\ref{Fig2}\,(B), demonstrate that the XY8 $^{1}$H NMR spectrum is contaminated by the presence of diamond $^{13}$C spins, while the Spin-Locking approach yields the same results independent of the amount of nearby $^{13}$C.
Together with the filter functions discussed above, these results demonstrate the immunity of Spin-Lock NMR to harmonic frequencies, highlighting its key advantage over XY8 in NMR-based applications.

% ---------------------------------------------------------
% NMR RESOLUTION
% ---------------------------------------------------------
The other potential weakness of XY8 is its limited spectral resolution. 
For hardware with a time resolution of $\Delta t\ll\tau_0$, the frequency resolution $\Delta f$ in XY8 NMR is
\begin{align}\label{eq:spectralresolution}
\Delta f &= \frac{1}{2\tau_0} -\frac{1}{2(\tau_0+\Delta t)}\\\nonumber
&\approx\frac{\Delta t}{2\tau_0^2} = 2\Delta t\,(\gamma_{\rm nuc}B_{\rm ext})^2\comma
\end{align} 
where we have used that on resonance, $1/(2\tau_0)=\gamma_{\rm nuc}B_{\rm ext}$ with $\gamma_{\rm nuc}$ being the gyromagnetic ratio of the target NMR species.
Even with state-of-the-art hardware that provides $\Delta t = \SI{0.5}{ns}$, Eq.~\eqref{eq:spectralresolution} predicts a spectral resolution for $^{1}$H NMR of just \SI{4.89}{kHz} at a magnetic field of $B_{\rm ext}=\SI{519.6}{G}$.
We confirm this result experimentally by measuring XY8 $^{1}$H NMR under these conditions (Fig~\ref{Fig2}\,(C)), where we obtain $\Delta f = \SI{4.93}{kHz}$, which leads to sparse sampling of the \SI{22.2}{kHz} (full-width-at-half-maximum) wide $^{1}$H NMR peak shown in Fig~\ref{Fig2}\,(C). 
As mentioned above, one could improve the spectral resolution by measuring at $B_{\rm ext}\ll\SI{500}{G}$, but that would lead to a loss of the nitrogen spin hyperpolarization and thereby to a reduction of the SNR.

The spectral resolution of Spin-Lock NMR, on the other hand, is limited only by the step size with which~$\Omega_R$, i.e. the applied microwave field amplitude, can be controlled experimentally. 
In Fig.~\ref{Fig2}\,(C), we present Spin-Lock NMR data taken under the same experimental conditions as the XY8 dataset, where we demonstrate a frequency resolution of \SI{0.30}{kHz}, which exceeds the resolution limit of XY8 by more than a factor of $10$.
In principle, one could achieve even higher frequency resolution with Spin-Lock NMR, technically down to the noise limit of the deployed electronic hardware.
We conclude that Spin-Lock NMR provides not only full immunity to $^{13}$C-related signals, but it also offers substantially higher spectral resolution at the experimentally advantageous bias field of \SI{500}{G}.
Achieving comparable resolution with XY8 would require sequence modifications with increased complexity, typically at the expense of SNR~\cite{Boss2017a,Jiang2023a}.

% ---------------------------------------------------------
% Section II
% ---------------------------------------------------------
\section{Depth Determination with Spin-Lock NMR}
\label{sec:depth_spinlock_nmr}

% ---------------------------------------------------------
% Rotating Frame Hamiltonian
% ---------------------------------------------------------
In this section, we sketch the derivation of the spin contrast model used for depth determination with Spin-Lock NMR.
The full derivation is presented in App.~\ref{app:spinlock_master_equation_derivation}, \ref{app:spinlock_meq_spin_contrast}, and \ref{app:correlationfunctions}.
We begin with the laboratory frame Hamiltonian,
\begin{align}\label{eq:main_spinlock_hamiltonian_lab}
    \hat H_{\mathrm{lab}}
    &=
    \hbar\omega_0 \hat S_z
    +2\hbar\Omega_R\cos(\omega_0 t)\hat{S}_x
    \\\nonumber &\qquad+\hbar\gamma_{\mathrm{NV}}\hat{\bm B}_{\rm N}\cdot\hat{\bm S}
    +\hbar\omega_L\sum_j \hat I_z^j \comma
\end{align}
where \(\hat S_k=\hat\sigma_k/2\) are spin-\(1/2\) operators (with the Pauli matrices $\sigma_k$) acting on the effective two-level system spanned by \(\ket{0}\) and \(\ket{-1}\), and the coordinate system $\{x,y,z\}$ is chosen such that the $z$-axis points along the NV symmetry axis.
The first term sets the energy splitting \(\hbar\omega_0\) between these two levels.
The second term describes the resonant Spin-Locking microwave field of Rabi frequency~$2\Omega_R$ (the factor 2 is introduced for ease of notation in the rotating frame below), where we assume the field to be linearly polarized, and without loss of generality, we let it point along the $x$-axis.
The third term describes the dipolar interaction of the NV with the nuclear spin bath, where \(\hat{\bm B}_{\rm N}\) is the operator that describes the magnetic field generated by the $^{1}$H nuclear spins at the NV position, and $\gamma_{\rm NV}=2\pi\times\SI{28}{GHz/T}$ is the NV gyromagnetic ratio.
Finally, the fourth term describes the energy splitting \(\hbar \omega_L\) of the $^{1}$H spins, enumerated with index~$j$.
To continue, we move to a rotating frame of the NV and nuclear spins.
In the rotating wave approximation with respect to $\omega_0 \gg \omega_L,\Omega_R$, this yields the Hamiltonian
\begin{align}
    \hat H = \hbar \Omega_R \hat S_x
    + \hbar \gamma_\mathrm{NV}\hat S_z \otimes \hat B(t)\comma
    \label{eq:main_spinlock_hamiltonian}
\end{align}
where \(\hat B(t)\) is the $^{1}$H-generated magnetic-field operator at the NV position in the rotating frame.
Notably, the only secular terms remaining are the ones that couple to the NV through $\hat S_z$.

% ---------------------------------------------------------
% Derivation of the QME
% ---------------------------------------------------------
We now treat the $^{1}$H ensemble as a fluctuating spin bath in a stationary state and derive a Markovian master equation for the reduced NV density matrix. 
This Markovian description is valid in the regime where the $^{1}$H correlation time $T_\mathrm{nuc}$ is short compared to the Spin-Lock relaxation time $T_{1\rho}$, i.e. where $T_\mathrm{nuc}/T_{1\rho}\ll1$.
For our experiments, this is fulfilled because $T_{\rm nuc}$ is a few tens of microseconds at room temperature~\cite{Pham2016a}, and $T_{1\rho}$ ranges from $\sim$\SI{100}{\micro s} to few ms.
We further assume that the $^{1}$H spin bath is initially uncorrelated with the NV and that it is effectively well described by an infinite-temperature, thermal state~\(\hat\rho_B\propto\hat\openone\).
Finally, we use the secular approximation in the dressed-state basis, which requires \(\Omega_R\approx2\pi\times\SI{2}{MHz}\) to be large compared to the relaxation rate $1/T_{1\rho}<\SI{10}{kHz}$.
Under these assumptions, as shown in App.~\ref{app:spinlock_master_equation_derivation}, the reduced NV density matrix $\hat{\rho}_{\rm NV}$ obeys the quantum master equation (QME)
\begin{multline}
    \frac{\text{d}}{\text{d}t}\hat\rho_\mathrm{NV}(t)
    ={} -i\Bigl[\Omega_R\hat S_x,\hat\rho_\mathrm{NV}(t)\Bigr] 
    \\
    +\Gamma_\downarrow\mathcal D\Bigl[\ket{-}\bra{+}\Bigr]\hat\rho_\mathrm{NV}(t)
    +\Gamma_\uparrow\mathcal D\Bigl[\ket{+}\bra{-}\Bigr]\hat\rho_\mathrm{NV}(t)\comma
    \label{eq:main_spinlock_master_equation}
\end{multline}
where
\begin{align}
    \mathcal D[\hat L]\,\hat\rho
    =\hat L\hat\rho\hat L^\dagger
    -\frac{1}{2}\left\{\hat L^\dagger\hat L,\hat\rho\right\}
\end{align}
is the Lindblad dissipator, and where the transition rates between the dressed states $\ket{\pm}$ are
\begin{align}
    \Gamma_\downarrow
    =\frac{\gamma_\mathrm{NV}^2}{4} S_B(+\Omega_R)\comma
    \qquad
    \Gamma_\uparrow
    =\frac{\gamma_\mathrm{NV}^2}{4} S_B(-\Omega_R)\fullstop
    \label{eq:main_spinlock_transition_rates}
\end{align}
Here, $S_B(\omega)$ is the (two-sided) magnetic noise spectral density
\begin{align}
    S_B(\omega)
    &=\int_{-\infty}^{+\infty} \text{d}t\,e^{i\omega t}\,\operatorname{Tr}_B\left\{\hat B(t)\hat B(0)\,\hat\rho_B\right\}\fullstop
    \label{eq:main_spinlock_spectrum_definition}
\end{align}
For the infinite-temperature nuclear spin bath we consider, $S_B(\omega)$ is symmetric, i.e. \(S_B(+\omega)= S_B(-\omega)\), and we therefore write
\begin{align}
    \Gamma_\downarrow+\Gamma_\uparrow
    =\frac{\gamma_\mathrm{NV}^2}{2}S_B(\Omega_R)
    :=\frac{1}{T_{1\rho}}
    \label{eq:main_spinlock_relaxation_rate_general}
\end{align}
in terms of a global Spin-Lock relaxation time~$T_{1\rho}$.

% ---------------------------------------------------------
% FIGURE 3
% ---------------------------------------------------------
\begin{figure*}[t]
\includegraphics{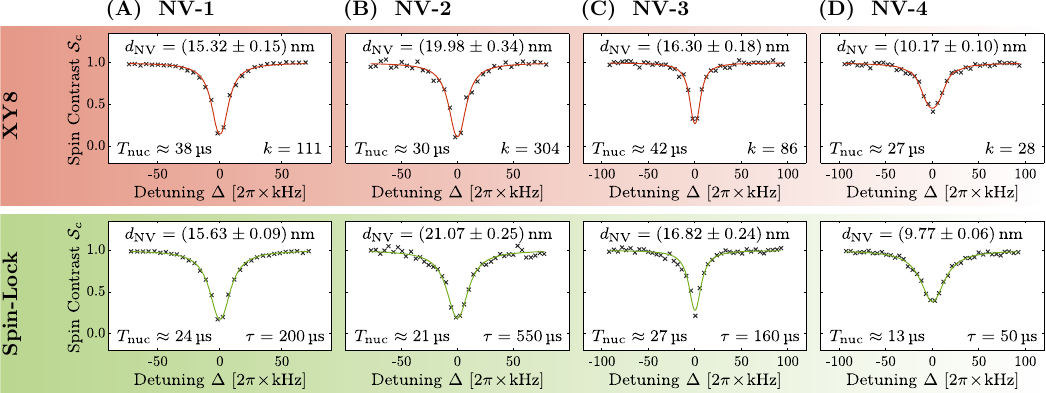}
\caption{\label{Fig3} 
\textbf{Comparison of XY8 and Spin-Lock depth determination on immersion oil $^{1}$H spins}.
We measure the $^{1}$H NMR signal of immersion oil on the diamond surface with both sequences on four different NVs.
These NVs, labeled NV-1 to NV-4, are located in an isotopically purified diamond crystal and were created by nitrogen ion implantation at an incidence angle of \SI{7}{\degree} and an energy of \SI{6}{keV}, resulting in a nominal implantation depth estimated by SRIM to be~\SI{9}{nm}.
For each NV, the Spin-Lock duration~$\tau$ is chosen to maximize the NMR SNR, and each XY8 measurement is performed with an XY8 repetition number $k$ such that the total XY8 sequence duration $8k\tau_0$ is equal to~$\tau$.
To extract the NV depth~$d_{\rm NV}$, we fit the XY8 data with the model of Pham \textit{et al.}~\cite{Pham2016a}, and the Spin-Lock data with the model derived here (cf.  Eq.~\eqref{eq:main_spinlock_final_fit_function}),  where in both cases, $d_{\rm NV}$ and $T_{\rm nuc}$ are the only free fit parameters.
We find excellent agreement of the determined depth values between XY8 and Spin-Lock, demonstrating the validity of the Spin-Lock based depth determination developed here.
The fitted values for $T_{\rm nuc}$ for the four NVs are, for XY8 
\SI{38.06\pm2.24}{\micro s},
\SI{30.01\pm2.73}{\micro s},
\SI{42.41\pm2.94}{\micro s}, and
\SI{26.93\pm2.09}{\micro s};
and for Spin-Lock
\SI{24.20\pm0.73}{\micro s},
\SI{20.56\pm1.22}{\micro s},
\SI{27.39\pm1.89}{\micro s}, and
\SI{12.84\pm0.38}{\micro s}.
Errors on $d_{\rm NV}$ and $T_{\rm nuc}$ are purely statistical, i.e. fit standard errors.
}
\end{figure*}

% ---------------------------------------------------------
% From QME to the dersired NMR Fit Function
% ---------------------------------------------------------
The quantity of interest to our work is the normalized spin contrast~$\mathcal{S}_c$, defined in Eq.~\eqref{eq:spin_contrast_def}, as a function of Spin-Lock pulse  duration~$\tau$ and Rabi frequency~$\Omega_R$.
In App.~\ref{app:spinlock_meq_spin_contrast}, we solve the QME~\eqref{eq:main_spinlock_master_equation} and find that $\mathcal{S}_c$ obeys a simple exponential decay characterized by $T_{1\rho}$:
\begin{align}
    \mathcal{S}_c =\exp\Biggl( -\frac{\tau}{T_{1\rho}} \Biggr)\fullstop
    \label{eq:main_spinlock_contrast_general}
\end{align}
To obtain a fit function that depends explicitly on the NV depth, we now evaluate the nuclear spin noise spectral density $S_B(\omega)$ from Eq.~\eqref{eq:main_spinlock_spectrum_definition} for a $^{1}$H ensemble on the diamond surface (cf. App.~\ref{app:correlationfunctions}).
For this, we model the $^{1}$H nuclei as non-interacting, infinite-temperature spin-1/2 magnetic dipoles with volume density \(\rho_V\), such that the nuclear spin correlation function is
\begin{equation}\label{eq:main_nuc_spin_correlator_with_decay}
{\rm Tr}_{\rm nuc}\qty{\hat I_\alpha^j(t)\hat I_\beta^i(0)\,\hat{\rho}_{B}}=\frac{1}{4}\,\delta_{ij}\delta_{\alpha\beta}\,e^{-|t|/T_{\rm nuc}}\comma
\end{equation}
where $\alpha,\beta\in\{x,y,z\}$ and where $i,j$ index individual nuclear spins. 
Following Ref.~\cite{Pham2016a}, we model the loss of correlation due to molecular diffusion through the NV sensing volume with an exponential decay with correlation time \(T_{\rm nuc}\).
%Importantly, and inspired by the approach taken in Ref.~\cite{Pham2016a}, we have included an exponential decay in Eq.~\eqref{eq:main_nuc_spin_correlator_with_decay} to account for a non-zero nuclear correlation time \(T_{\rm nuc}\) due to diffusion of nuclear spins in and out of the NV sensing volume.
With this nuclear spin correlation function, we evaluate Eq.~\eqref{eq:main_spinlock_spectrum_definition} and find
\begin{equation}
S_B(\omega) = C_0L(\omega)+C_{\rm NMR}\biggl(L(\omega-\omega_L)+L(\omega+\omega_L)\biggr)\comma
\label{eq:main_spinlock_nmr_components_to_SB}
\end{equation}
where $C_0$ and $C_{\rm NMR}$ are constants that depend on the NV depth and the geometry of the nuclear spin bath, and where
\begin{align}
L(\omega)
&=
\frac{T_{\rm nuc}}{1+\omega^2T_{\rm nuc}^2}.
\label{eq:main_spinlock_nmr_lorentzian}
\end{align}
is a Lorentzian function with a width set by $1/T_{\rm nuc}$.
According to Eq.~\eqref{eq:main_spinlock_relaxation_rate_general}, we are only interested in evaluating $S_B(\omega)$ at $\omega=\Omega_R$.
Since we operate near the HH resonance condition where $\Omega_R\approx\omega_L$, the resonant term $L(\omega-\omega_L)$ dominates Eq.~\eqref{eq:main_spinlock_nmr_components_to_SB}, such that the zero-frequency peak $L(\omega)$ and the off-resonant counter-rotating contribution $L(\omega+\omega_L)$ can both be neglected.
Introducing the NMR detuning $\Delta=\omega-\omega_L$, we therefore write the relevant NMR spectrum as
\begin{align}
S_B(\Delta)
&=
C_{\rm NMR}L(\Delta)\fullstop
\label{eq:main_spinlock_nmr_spectrum}
\end{align}
Finally, we evaluate $C_{\rm NMR}$ by solving the geometric dipolar integral over the nuclear spin ensemble volume.
As shown in App.~\ref{app:correlationfunctions}, for a semi-infinite $^{1}$H-containing layer above a [100]-oriented diamond surface, this integral yields
\begin{align}\label{eq:main_spinlock_Cnmr_volume}
C_{\rm NMR}
=
\frac{5\rho_V(\hbar\mu_0\gamma_{\rm H})^2}
{1536\pi\,d_{\rm NV}^3}\comma
\end{align}
where $d_{\rm NV}$ is the depth of the NV center, $\rho_V$ is the volume spin density of the nuclear ensemble, $\gamma_H$ is the gyromagnetic ratio of the $^{1}$H nuclei, and $\mu_0$ is the vacuum permeability.
With this, the final Spin-Lock NMR fit function becomes
\begin{align}\label{eq:main_spinlock_final_fit_function}
\mathcal{S}_c &=\exp\Biggl( -\frac{\tau}{T_{1\rho}} \Biggr)\\\nonumber
&= \exp\Biggl( -\frac{1}{2}\gamma_{\rm NV}^2\tau\,C_{\rm NMR}L(\Omega_R-\omega_L)\Biggr)\\\nonumber
&= \exp\Biggl(-\frac{1}{2}\gamma_{\rm NV}^2\tau\,\frac{5\,\rho_V(\hbar\mu_0\gamma_{\rm H})^2}{1536\,\pi\cdot d_{\rm NV}^3}\cdot\frac{T_{\rm nuc}}{1+\Delta^2 T_{\rm nuc}^2}\Biggr)~.
\end{align}

% ---------------------------------------------------------
% FIGURE 4
% ---------------------------------------------------------
\begin{figure*}[t]
\includegraphics{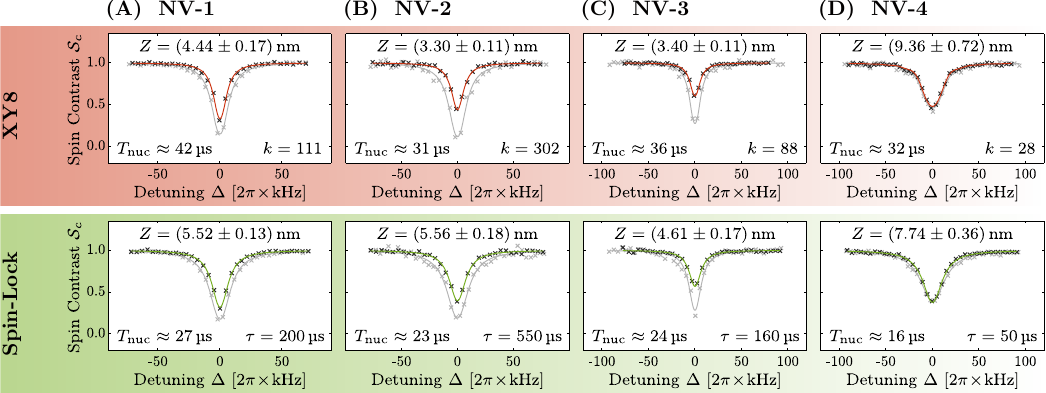}
\caption{\label{Fig4} 
\textbf{Analysis of an adsorbate surface layer containing $^{1}$H spins}.
NMR data taken on the same NVs as in Fig.~\ref{Fig3}, acquired before the immersion oil was applied, i.e. the measurements were performed with a clean (oxygen-terminated) diamond surface.
The resulting data taken by XY8 and Spin-Lock sequences are shown in green and red respectively, and the immersion oil data from Fig.~\ref{Fig3} are shown in grey for comparison.
Clearly, even in the absence of oil, all four NVs still show an $^1$H NMR signature, which is commonly assigned to a $\sim$\SI{1}{nm} thick, $^1$H-containing adsorbate layer on the diamond surface, as previously reported in~\cite{Abendroth2022a,Xu2025a,Xu2026a,DeVience2015a,Loretz2014a,Rugar2015a,Zheng2026a,Bruckmaier2023a}.
Similar to the approach taken in Refs.\,\cite{Xu2025a} and \cite{Abendroth2022a}, we fit the thickness~$Z$ of the adsorbate layer with the two NMR models by fixing the NV depth to the values obtained in Fig.~\ref{Fig3}, leaving $Z$ and $T_{\rm nuc}$ as free parameters.
Assuming an adsorbate layer $^1$H spin density equal to that of liquid water, $\rho_V=\SI{67}{nm^{-3}}$, we obtain layer thicknesses of a few nanometers, which is more than the $\sim$\SI{1}{nm} thick layers reported in other work~\cite{Abendroth2022a,Xu2025a,Xu2026a,DeVience2015a,Loretz2014a,Rugar2015a,Zheng2026a,Bruckmaier2023a}.
These results indicate that water alone is not sufficient to explain our data, but that there is an additional proton source, such as a dense hydrocarbon layer or $^{1}$H incorporated into the near-surface diamond.
}
\end{figure*}

% ---------------------------------------------------------
% NMR Data for Depth-Fit Comparison
% ---------------------------------------------------------
Next, we use Eq.~\eqref{eq:main_spinlock_final_fit_function} to fit the depths of individual NV centers via Spin-Lock NMR, and then compare the results with those of the established XY8-based depth determination protocol.
To that end, immersion oil (Leica Type F, ISO 8036) is applied to an isotopically purified diamond with a negligible $^{13}$C concentration, and $^{1}$H NMR is measured with both sequences on four single NV centers.
These NVs, labeled NV-1 to NV-4, were created by nitrogen ion implantation at \SI{6}{keV} and an incident angle of \SI{7}{\degree}~(see App.~\ref{app:diamondsamples}), resulting in a nominal implantation depth estimated by SRIM to be~\SI{9}{nm}.
For each NV, the Spin-Lock NMR is recorded with a Spin-Lock duration~$\tau$ that maximizes the SNR of the NMR dip.
This optimal value of~$\tau$, typically on the order of $\sim$$0.1\cdot T_{1\rho}$ to $\sim$$0.8\cdot T_{1\rho}$, is determined beforehand by measuring Spin-Lock NMR for different~$\tau$, as described in App.~\ref{app:snr_optimization}.
The XY8 measurements are then performed with a repetition number~$k$ (see Fig.~\ref{Fig1}\,(B)) chosen such that the total XY8 sequence duration~$8k\tau_0$ matches the corresponding Spin-Lock duration~$\tau$.
Furthermore, the XY8 spectra are recorded at the spectral resolution limit set by Eq.~\eqref{eq:spectralresolution}, and the Spin-Lock measurements are configured to provide the same spectral resolution for a direct comparison.
The resulting SNR-optimized NMR spectra are shown in Fig.~\ref{Fig3}.

% ---------------------------------------------------------
% Discussion of Fitted Depths
% ---------------------------------------------------------
To extract the NV depth~$d_{\rm NV}$, the XY8 spectra are fitted using the model by Pham \textit{\textit{et al.}}~\cite{Pham2016a}, while the Spin-Lock spectra are fitted with Eq.~\eqref{eq:main_spinlock_final_fit_function}.
In both cases, the free fit parameters are $d_{\rm NV}$ and $T_{\rm nuc}$, while the $^{1}$H spin density of the immersion oil is fixed to $\rho_V=\SI{68}{nm^{-3}}$~\cite{Pham2016a}.
The extracted NV depths are listed in Fig.~\ref{Fig3}, where the error bars denote the standard errors of the non-linear regression coefficients.
The fitted depth values range from 10 to \SI{20}{nm}, which is in good agreement with the nominal nitrogen implantation depth of \SI{9}{nm} predicted by SRIM for this diamond (taking into account that SRIM neglects crystal channeling~\cite{Lehtinen2016a} and thus underestimates the depth by a factor of up to two~\cite{Toyli2010a}).
Most importantly, for each of the four investigated NV centers, we find excellent agreement of the depth fitted by Spin-Lock and XY8 (with relative differences of $<$\SI{3}{\percent}), demonstrating that our Spin-Lock approach reproduces the results of the established depth determination scheme.

% ---------------------------------------------------------
% Discussion of Fitted Tnuc
% ---------------------------------------------------------
While the fitted NV depths agree well between Spin-Lock and XY8, the extracted values of $T_{\rm nuc}$ differ systematically, with the Spin-Lock fits yielding values up to two times smaller than those obtained from XY8.
This discrepancy is not a deficiency in either theoretical model, but it instead reflects accurately that the obtained Spin-Lock NMR dips are indeed slightly broader than the corresponding XY8 dips.
We attribute this broadening to fluctuations in the microwave power caused by temperature drifts and heating of microwave components such as the loop antenna.
Since the Spin-Lock detection frequency is set by the Rabi frequency, $f=\Omega_R/(2\pi)$, such microwave power fluctuations shift the NMR resonance, leading to an effective broadening of the measured spectrum when averaged over time.
To suppress the effect of slow drifts, all NMR data in this work are acquired in one-hour-long measurements with recalibration between runs before carefully averaging into a single multi-hour spectrum (see App.~\ref{app:data_normalization}).
Power fluctuations on timescales shorter than one hour, however, still broaden the Spin-Lock resonance.
In contrast, the XY8 detection frequency $f=1/(2\tau_0)$ is independent of the microwave power, and the sequence is inherently robust against small pulse errors caused by microwave power fluctuations~\cite{Lee1987a,Souza2011a}.
For these reasons, we expect the XY8 spectra to be largely unaffected by the broadening mechanism present in Spin-Lock NMR, which explains the difference in dip width and fitted $T_{\rm nuc}$ between the two approaches.
Most importantly, the fitted depth~$d_{\rm NV}$ is not affected by this power fluctuation induced broadening.
This is due to the NV depth being set by the overall area of the NMR dip, which remains unchanged by the described broadening mechanism.
For the four NVs studied in Fig.~\ref{Fig3}, said area is indeed nearly identical between XY8 and Spin-Lock, with a relative difference of $<$\SI{3}{\percent}.

% ---------------------------------------------------------
% Section: Adsorbate Surface Layer
% ---------------------------------------------------------
\section{Analysis of a 1H-containing adsorbate surface layer}
\label{sec:adsorbate_layer}

As an application of the depth determination technique presented above, we investigate a $^{1}$H NMR signal that we observe on all of our diamond samples in the absence of immersion oil, i.e. with a clean (oxygen-terminated) diamond surface.
Similar ubiquitous $^{1}$H signals have been reported in previous NV-NMR experiments~\cite{Abendroth2022a,Xu2025a,Xu2026a,DeVience2015a,Loretz2014a,Rugar2015a,Zheng2026a,Bruckmaier2023a} and are commonly attributed to an approximately \SI{1}{nm}-thick layer of adsorbed hydrocarbons or immobilized, solid-like water.
This interpretation is supported by recent studies showing that molecular diffusion is strongly suppressed in two-dimensionally nano-confined water with $<$\SI{2}{nm} thickness, making water $^{1}$H NMR feasible~\cite{Zheng2026a,Bruckmaier2023a,Li2024a,Staudacher2015a}, in contrast to unconfined water where rapid diffusion 
%($D\approx\SI{2000}{nm^2/\micro s}$~\cite{Holz2000a,Mills1973a})
broadens the NMR linewidth to more than \SI{1}{MHz}.

To characterize this surface layer, we record XY8 and Spin-Lock NMR spectra on the same four NV centers as shown in Fig.~\ref{Fig3}, but without applying immersion oil~(Fig.~\ref{Fig4}).
We fit the obtained spectra with both depth determination approaches using 
\begin{align}\label{eq:main:cnmrforfinitelayers}
C_{\rm NMR}^{\rm layer} &= \frac{5\,\rho_V(\hbar\mu_0\gamma_{\rm H})^2}{1536\,\pi}\left(\frac{1}{d_{\rm NV}^3}-\frac{1}{(d_{\rm NV}+Z)^3}\right)\comma
\end{align}
instead of Eq.~\eqref{eq:main_spinlock_Cnmr_volume} to account for the finite proton layer thickness~$Z$ (see App.~\ref{app:correlationfunctions}). 
The proton spin density is assumed to be equal to that of water, $\rho_V\approx\SI{67}{nm^{-3}}$, and the NV depth $d_{\rm NV}$ is fixed to the value determined for the respective NV in Fig.~\ref{Fig3}, leaving only $T_{\rm nuc}$ and $Z$ as free fitting parameters.
The resulting layer thicknesses range from 3.3 to \SI{9.4}{nm}, substantially exceeding the $\sim$\SI{1}{nm} thickness reported by others~\cite{Abendroth2022a,Xu2025a,Xu2026a,DeVience2015a,Loretz2014a,Rugar2015a,Zheng2026a,Bruckmaier2023a}.
At such thicknesses, water diffusion is expected to approach its bulk value, making water alone an unlikely explanation for the observed NMR signal.
Instead, our results suggest an additional proton source, such as a dense hydrocarbon layer or $^{1}$H spins incorporated into the near-surface diamond material.

Our measurements do not establish whether the application of immersion oil removes (some of) the native adsorbates~\cite{DeVience2015a,Xu2026a}, or whether it instead forms an additional layer above them.
Regardless of the microscopic picture, the immersion-oil-based depth determination protocol discussed in this work remains essentially unaffected by a thin adsorbate layer, because the proton spin density of water (\SI{67}{nm^{-3}}) is nearly identical to that of immersion oil (\SI{68}{nm^{-3}}).
In addition, the inferred NV depth is only weakly sensitive to moderate uncertainties in the assumed adsorbate proton density, because for a semi-infinite proton bath, the signal amplitude depends on the combination $\rho_V/d_{\rm NV}^{3}$ (Eq.~\eqref{eq:main_spinlock_Cnmr_volume}), such that the fitted depth scales as $d_{\rm NV}\propto(\rho_V)^{1/3}$.
A water-like adsorbate layer is therefore not expected to produce a significant bias in the immersion-oil-based depth determination, irrespective of whether it is displaced by the oil or remains underneath it.
By contrast, a substantially different proton density than that of water or an appreciable contribution from subsurface $^{1}$H could shift the inferred depth and should be regarded as a residual systematic uncertainty.
Determining the microscopic origin of the ubiquitous $^{1}$H signal therefore remains relevant for further improving the accuracy of NV-depth measurements.

% ---------------------------------------------------------
% Section: Conclusion
% ---------------------------------------------------------
\section*{Conclusion}

% ---------------------------------------------------------
% Paper Summary
% ---------------------------------------------------------
We have introduced Spin-Lock as an alternative to XY8 for determining the depth of shallow NV centers in diamond from the $^{1}$H NMR signal of immersion oil deposited on the diamond surface.
Experimentally, we have demonstrated that unlike XY8, Spin-Lock is insensitive to harmonic and subharmonic spectral ambiguities while providing substantially higher frequency resolution.
To enable quantitative depth determination, we have derived a Markovian spin-contrast model that directly relates the measured Spin-Lock NMR spectrum to the NV depth.
Applying this model to four individual NV centers yields depth estimates in excellent agreement with the established XY8-based protocol, with relative differences of $<$\SI{3}{\percent}.
We further find that a water-like $^{1}$H-containing adsorbate layer is not expected to introduce a significant bias in the fitted depth because its proton density is close to that of the immersion oil.
Our results establish Spin-Lock NMR as a robust and quantitative alternative for single-NV depth determination.

% ---------------------------------------------------------
% Outlook
% ---------------------------------------------------------
In its current form, our Spin-Lock technique is particularly well suited to address applications where harmonic contamination, high magnetic fields, or narrow NMR linewidths limit the performance of pulsed dynamical decoupling.
Our protocol's main weaknesses are its susceptibility to microwave power fluctuations which broaden the measured NMR resonance, and the need to calibrate the NMR frequency axis through Rabi measurements.
Future improvements should therefore focus on stabilizing or actively tracking the Rabi frequency during long measurements to suppress power-induced linewidth broadening and thereby improve access to the intrinsic nuclear correlation time $T_{\rm nuc}$.
In addition, the microwave power calibration could be eliminated altogether by applying two auxiliary microwave tones detuned from the expected nuclear Larmor frequency, by e.g. a few \SI{100}{kHz}, to produce two reference resonances at well-known frequencies to directly calibrate the NMR frequency axis.

% ---------------------------------------------------------
% Acknowledgments
% ---------------------------------------------------------
\begin{acknowledgments}
A.D. acknowledges financial support by the QCQT PhD school.
A.D. and P.P. acknowledge financial support by the Swiss National Science Foundation (Eccellenza Professorial Fellowship PCEFP2\_194268).  
B.B. and P.M. acknowledge financial support by the SERI Swiss quantum call project “QuantumLeap” and by the SNSF quantum transitional call project “ensQsens”.
\end{acknowledgments}

% ---------------------------------------------------------
% Appendix
% ---------------------------------------------------------
\appendix 
\section{Master-equation derivation}
\label{app:spinlock_master_equation_derivation}

In this appendix we derive a master-equation description of the Spin-Lock NMR signal following the procedure outlined in Ref.~\cite{SciPostPhysLectNotes.126}.
The goal is to describe the decay of the Spin-Locked NV state due to the fluctuating magnetic field generated by the $^{1}$H spin bath.
We treat the $^{1}$H ensemble as a fluctuating bath and derive a Markovian master equation for the reduced NV state.
This derivation fixes the relation between the measured spin contrast and the magnetic noise spectrum at the dressed-state transition frequency.

\subsection{Rotating-frame Hamiltonian}
\label{app:spinlock_meq_rotating_frame}

We start from the lab-frame Hamiltonian
\begin{align} \label{eq:spinlock_meq_lab_hamiltonian}
    \hat H_{\mathrm{lab}}
    &=
    \hbar\omega_0 \hat S_z
    +2\hbar\Omega_R\cos(\omega_0 t)\hat{S}_x
    \\\nonumber &\qquad+\hbar\gamma_{\mathrm{NV}}\hat{\bm B}_{\rm N}\cdot\hat{\bm S}
    +\hbar\omega_L\sum_j \hat I_z^j \fullstop
\end{align}
Here, the first term describes the transition frequency \(\omega_0\) of the effective NV two-level system spanned by $\ket{0}$ and $\ket{-1}$.
The second term describes the resonant Spin-Locking microwave field of Rabi frequency~$2\Omega_R$ (the factor 2 is introduced for ease of notation in the rotating frame below), where we assume the field to be linearly polarized, and without loss of generality, we let it point along the NV $x$-axis.
The third term is the interaction of the NV with the nuclear spin bath, where \(\hat{\bm B}_{\rm N}\) is the operator that describes the magnetic field generated by the $^{1}$H nuclear spins at the NV position.
Lastly, the fourth term describes the $^{1}$H spins that precess at the Larmor frequency \(\omega_L\).
We use spin-\(1/2\) operators for the NV center, such that \(\hat S_k=\hat\sigma_k/2\), and we assume that the states \(\ket{0}\) and \(\ket{-1}\) form an isolated two-level system.

We now transform to the frame rotating with the NV transition frequency and with the $^{1}$H Larmor frequency, using the unitary operator
\begin{equation}\label{app:rotatingframeunitaryoperator}
    \hat{U}(t) = \exp\Bigl(i \omega_0 t \cdot \hat{S}_z + i \omega_L t \sum_j \hat{I}^{j}_{z} \Bigr)\fullstop
\end{equation}
After performing this rotation and applying the rotating-wave approximation to the microwave drive and to the NV-nuclei coupling, we obtain the Hamiltonian
\begin{align}
    \label{eq:spinlock_meq_rotating_hamiltonian}
    \hat H
    &= \hat{U}(t)\hat{H}_{\rm lab}\hat{U}^\dagger(t) + i\hbar\,\frac{\partial\hat{U}(t)}{\partial t}\,\hat{U}^\dagger(t)\\\nonumber
    &=
    \hbar\Omega_R \hat S_x
    +
    \hbar\gamma_{\mathrm{NV}}\hat S_z\otimes \hat B(t)\fullstop
\end{align}
In this equation, \(\Omega_R\) is the Rabi frequency of the Spin-Lock field, and \(\hat B(t)\) is the component of the $^{1}$H-generated magnetic-field operator that couples to the NV in the rotating frame.
We compute $\hat{B}(t)$ explicitly in App.~\ref{app:correlationfunctions} and continue here with the general variable~$\hat{B}(t)$.

Equation~\eqref{eq:spinlock_meq_rotating_hamiltonian} is the starting point for the master-equation treatment.
We split it into a system Hamiltonian and an interaction Hamiltonian,
\begin{align}
    \hat H_S
    =
    \hbar\Omega_R \hat S_x,
    \qquad
    \hat V(t)
    =
    \hbar\gamma_{\mathrm{NV}}\hat S_z\otimes \hat B(t).
    \label{eq:spinlock_meq_system_interaction_split}
\end{align}
The drive now defines the dressed-state energy splitting and the eigenstates of \(\hat S_x\) are
\begin{align}
    \ket{\pm}
    =
    \frac{1}{\sqrt{2}}\left(\ket{0} \pm \ket{-1}\right),
    \label{eq:spinlock_meq_pm_states}
\end{align}
with
\begin{align}
    \hat S_x\ket{\pm}
    =
    \pm\frac{1}{2}\ket{\pm}.
\end{align}
To express the coupling in this basis, we define the dressed-state transition operators
\begin{align}
    \hat S_+^{(x)}
    =
    \ket{+}\bra{-},
    \qquad
    \hat S_-^{(x)}
    =
    \ket{-}\bra{+}.
    \label{eq:spinlock_meq_x_ladder_ops}
\end{align}
The operator \(\hat S_z\) can be written as
\begin{align}
    \hat S_z
    =
    \frac{1}{2}
    \left(
    \hat S_+^{(x)}
    +
    \hat S_-^{(x)}
    \right).
    \label{eq:spinlock_meq_Sz_decomposition}
\end{align}
Thus, the interaction Hamiltonian becomes
\begin{align}
    \hat V(t)
    =
    \frac{\hbar\gamma_{\mathrm{NV}}}{2}
    \left(
    \hat S_+^{(x)}
    +
    \hat S_-^{(x)}
    \right)
    \otimes \hat B(t).
    \label{eq:spinlock_meq_interaction_ladder_form}
\end{align}

\subsection{Born--Markov equation}
\label{app:spinlock_meq_born_markov_section}

The total density matrix is denoted by \(\hat\rho_{\mathrm{tot}}\), and the reduced NV density matrix is
\begin{align}
    \hat\rho_{\mathrm{NV}}(t)
    =
    \operatorname{Tr}_{\mathrm{nuc}}\left\{\hat\rho_{\mathrm{tot}}(t)\right\}.
\end{align}
We assume that the initial state factorizes as
\begin{align}
    \hat\rho_{\mathrm{tot}}(0)
    =
    \hat\rho_{\mathrm{NV}}(0)\otimes \hat\rho_B,
    \label{eq:spinlock_meq_factorized_initial_state}
\end{align}
where $\hat\rho_B$ is the stationary state of the $^{1}$H bath.
For the present application we take this state to be the infinite-temperature state,
\begin{align}
    \hat\rho_B
    =
    \bigotimes_{j=1}^N \frac{\openone^j}{2},
    \label{eq:spinlock_meq_infinite_temperature_bath}
\end{align}
where \(N\) is the number of $^{1}$H spins.
This directly implies that we have no net magnetization in the $^{1}$H bath, 
\begin{align}
    \operatorname{Tr}_{B}\left\{\hat B(t)\hat\rho_{B}\right\}
    =
    0.
    \label{eq:spinlock_meq_zero_bath_mean}
\end{align}

We move to the interaction picture with respect to the Spin-Lock Hamiltonian \(\hat H_S\).
The corresponding transformation is generated by
\begin{align}
    \hat U_x(t)
    =
    \exp\left(i\Omega_R t\hat S_x\right).
    \label{eq:spinlock_meq_interaction_picture_unitary}
\end{align}
The interaction-picture interaction Hamiltonian is
\begin{align}
    \tilde V(t)
    =
    \hat U_x(t)\hat V(t)\hat U_x^\dagger(t).
\end{align}
Using Eq.~\eqref{eq:spinlock_meq_x_ladder_ops}, we obtain
\begin{align}
    \hat U_x(t)\hat S_+^{(x)}\hat U_x^\dagger(t)
    &=
    e^{+i\Omega_R t}\hat S_+^{(x)},
    \\
    \hat U_x(t)\hat S_-^{(x)}\hat U_x^\dagger(t)
    &=
    e^{-i\Omega_R t}\hat S_-^{(x)}.
\end{align}
Hence
\begin{align}
    \tilde V(t)
    =
    \frac{\hbar\gamma_{\mathrm{NV}}}{2}
    \left(
    e^{+i\Omega_R t}\hat S_+^{(x)}
    +
    e^{-i\Omega_R t}\hat S_-^{(x)}
    \right)
    \otimes \hat B(t).
    \label{eq:spinlock_meq_interaction_picture_V}
\end{align}

The interaction-picture total density matrix now evolves according to
\begin{align}
    \frac{\text{d}}{\text{d}t}\tilde\rho_{\mathrm{tot}}(t)
    =
    -\frac{i}{\hbar}
    \left[
    \tilde V(t),
    \tilde\rho_{\mathrm{tot}}(t)
    \right].
    \label{eq:spinlock_meq_interaction_picture_eom}
\end{align}
Applying the weak-coupling and Markov approximations gives the Born--Markov equation
\begin{multline}
    \frac{\text{d}}{\text{d}t}\tilde\rho_{\mathrm{NV}}(t)
    =\\ 
    -
    \frac{1}{\hbar^2}
    \int_0^\infty \text{d}s\,
    \operatorname{Tr}_{\mathrm{nuc}}
    \Bigl\{
    \left[
    \tilde V(t),
    \left[
    \tilde V(t-s),
    \tilde\rho_{\mathrm{NV}}(t)\otimes \hat\rho_{B}
    \right]
    \right]
    \Bigr\}.
    \label{eq:spinlock_meq_born_markov}
\end{multline}
This equation is valid when the bath correlation time is short compared to the time scale on which \(\tilde\rho_{\mathrm{NV}}(t)\) changes~\cite{SciPostPhysLectNotes.126}.
In the present context, this means that the $^{1}$H correlation time is assumed to be short compared to the Spin-Lock relaxation time.

To evaluate Eq.~\eqref{eq:spinlock_meq_born_markov}, we introduce the bath correlation function
\begin{align}
    C_B(s)
    =
    \operatorname{Tr}_{B}
    \left\{
    \hat B(t)\hat B(t-s)\hat\rho_{B}
    \right\}.
    \label{eq:spinlock_meq_bath_correlation_definition}
\end{align}
Stationarity of the bath was used to write this correlation function as a function of the time difference \(s\).
Equivalently, one may set the reference time to zero and write \(C_B(s)=\operatorname{Tr}_{\mathrm{nuc}}\{\hat B(s)\hat B(0)\hat\tau_{\mathrm{nuc}}\}\).
The one-sided transform of the bath correlation function is defined as
\begin{align}
    \Gamma_B(\omega)
    =
    \int_0^\infty \text{d}s\,e^{i\omega s}C_B(s).
    \label{eq:spinlock_meq_one_sided_transform}
\end{align}
We decompose it into real and imaginary parts,
\begin{align}
    \Gamma_B(\omega)
    =
    \frac{1}{2}S_B(\omega)
    +
    i\Delta_B(\omega),
    \label{eq:spinlock_meq_gamma_decomposition}
\end{align}
where
\begin{align}
    S_B(\omega)
    =
    \int_{-\infty}^{+\infty}\text{d}s\, e^{i\omega s}C_B(s)
    \label{eq:spinlock_meq_two_sided_spectrum}
\end{align}
is the two-sided magnetic noise spectrum, and \(\Delta_B(\omega)\) gives the Lamb-shift contribution.

It is useful to write Eq.~\eqref{eq:spinlock_meq_interaction_picture_V} in terms of Bohr-frequency components.
We define
\begin{align}
    \hat A_{+\Omega_R}
    =
    \frac{\gamma_{\mathrm{NV}}}{2}\hat S_-^{(x)},
    \qquad
    \hat A_{-\Omega_R}
    =
    \frac{\gamma_{\mathrm{NV}}}{2}\hat S_+^{(x)}.
    \label{eq:spinlock_meq_Aomega_definition}
\end{align}
These operators obey
\begin{align}
    \left[
    \hat A_{+\Omega_R},
    \Omega_R\hat S_x
    \right]
    &=
    +\Omega_R \hat A_{+\Omega_R},
    \\
    \left[
    \hat A_{-\Omega_R},
    \Omega_R\hat S_x
    \right]
    &=
    -\Omega_R \hat A_{-\Omega_R}.
\end{align}
Thus \(\hat A_{+\Omega_R}\) removes the energy \(\hbar\Omega_R\) from the Spin-Locked NV, while \(\hat A_{-\Omega_R}\) adds the energy \(\hbar\Omega_R\).
With this notation,
\begin{align}
    \tilde V(t)
    =
    \hbar
    \sum_{\omega=\pm\Omega_R}
    e^{-i\omega t}\hat A_\omega\otimes \hat B(t).
    \label{eq:spinlock_meq_V_bohr_decomposition}
\end{align}
Substituting Eq.~\eqref{eq:spinlock_meq_V_bohr_decomposition} into Eq.~\eqref{eq:spinlock_meq_born_markov} gives
\begin{align}
    \frac{\text{d}}{\text{d}t}\,{\tilde\rho}_{\mathrm{NV}}
    ={}&
    \sum_{\omega,\omega'=\pm\Omega_R}
    e^{i(\omega-\omega')t}
    \Gamma_B(\omega')
    \nonumber\\
    &\times
    \left(
    \hat A_{\omega'}\tilde\rho_{\mathrm{NV}}\hat A_\omega^\dagger
    -
    \hat A_\omega^\dagger\hat A_{\omega'}\tilde\rho_{\mathrm{NV}}
    \right)
    +
    \mathrm{H.c.}
    \label{eq:spinlock_meq_presecular_redfield}
\end{align}
This is the Redfield equation in the Spin-Lock interaction picture \cite{REDFIELD19651}.
The terms with \(\omega=\omega'\) are time independent, whereas the terms with \(\omega\neq\omega'\) oscillate at the frequency
\begin{align}
    |\omega-\omega'|
    =
    2\Omega_R .
\end{align}
We now apply the secular approximation.
This approximation is valid when these oscillations are fast compared to the relaxation dynamics of the reduced NV state \cite{SciPostPhysLectNotes.126},
\begin{align}
    2\Omega_R T_{1\rho}
    \gg
    1.
    \label{eq:spinlock_meq_secular_condition}
\end{align}
This condition is fulfilled in the weak-coupling Spin-Lock regime, where the dressed-state splitting is much larger than the relaxation rate.
Under this approximation, we keep only the terms with \(\omega=\omega'\).
Equation~\eqref{eq:spinlock_meq_presecular_redfield} then becomes
\begin{multline}
    \frac{\text{d}}{\text{d}t}\tilde\rho_{\mathrm{NV}}(t)
    =
    -i
    \left[
    \hat H_{\mathrm{LS}},
    \tilde\rho_{\mathrm{NV}}(t)
    \right]
    \\ +
    \sum_{\omega=\pm\Omega_R}
    S_B(\omega)
    \mathcal D[\hat A_\omega]\tilde\rho_{\mathrm{NV}}(t),
    \label{eq:spinlock_meq_gkls_interaction_picture}
\end{multline}
where
\begin{align}
    \mathcal D[\hat L]\hat\rho
    =
    \hat L\hat\rho\hat L^\dagger
    -
    \frac{1}{2}
    \left\{
    \hat L^\dagger\hat L,
    \hat\rho
    \right\}
    \label{eq:spinlock_meq_dissipator}
\end{align}
and
\begin{align}
    \hat H_{\mathrm{LS}}
    =
    \sum_{\omega=\pm\Omega_R}
    \Delta_B(\omega)\hat A_\omega^\dagger\hat A_\omega
    \label{eq:spinlock_meq_lamb_shift}
\end{align}
is the Lamb-shift Hamiltonian.
Returning to the Spin-Lock rotating frame gives our final QME
\begin{align}
    \frac{\text{d}}{\text{d}t}\hat\rho_{\mathrm{NV}}(t)
    ={}&
    -i
    \left[
    \Omega_R\hat S_x+
    \hat H_{\mathrm{LS}},
    \hat\rho_{\mathrm{NV}}(t)
    \right] \label{eq:spinlock_meq_final_gkls}
    \\
    &+
    \Gamma_\downarrow
    \mathcal D[\hat S_-^{(x)}]\hat\rho_{\mathrm{NV}}(t)
    +
    \Gamma_\uparrow
    \mathcal D[\hat S_+^{(x)}]\hat\rho_{\mathrm{NV}}(t), \nonumber  
\end{align}
with the transition rates
\begin{align}
    \Gamma_\downarrow
    =
    \frac{\gamma_{\mathrm{NV}}^2}{4}
    S_B(+\Omega_R),
    \qquad
    \Gamma_\uparrow
    =
    \frac{\gamma_{\mathrm{NV}}^2}{4}
    S_B(-\Omega_R).
    \label{eq:spinlock_meq_transition_rates}
\end{align}
Here, \(\Gamma_\downarrow\) is the transition rate from \(\ket{+}\) to \(\ket{-}\), and \(\Gamma_\uparrow\) is the reverse transition rate.

\section{Spin contrast from the master equation}
\label{app:spinlock_meq_spin_contrast}

We can use the master equation to obtain the spin contrast at the end of the Spin-Lock pulse. The relevant observable is the dressed-state polarization along the Spin-Lock axis. We define the projectors
\begin{align}
    \hat P_+
    =
    \ket{+}\bra{+},
    \qquad
    \hat P_-
    =
    \ket{-}\bra{-},
\end{align}
and the corresponding probabilities
\begin{align}
    P_{\ket{+}}(t)
    &=
    \operatorname{Tr}\left\{\hat P_+\hat\rho_{\mathrm{NV}}(t)\right\},
    \qquad \\ 
    P_{\ket{-}}(t)
    &=
    \operatorname{Tr}\left\{\hat P_-\hat\rho_{\mathrm{NV}}(t)\right\}.
\end{align}
The spin contrast is the population difference in this basis,
\begin{align}
    \mathcal{S}_c(t)
    =
    P_{\ket{+}}(t)-P_{\ket{-}}(t)~.
    \label{eq:spinlock_meq_contrast_definition}
\end{align}
Equivalently, since \(\hat S_x\ket{\pm}=\pm\ket{\pm}/2\), this can be written as
\begin{align}
    \mathcal{S}_c(t)
    =
    2\operatorname{Tr}\left\{\hat S_x\hat\rho_{\mathrm{NV}}(t)\right\}.
    \label{eq:spinlock_meq_contrast_Sx}
\end{align}

To obtain the differential equation for \(\mathcal{S}_c(t)\), we first derive the equations of motion for the dressed-state populations.
The coherent part of Eq.~\eqref{eq:spinlock_meq_final_gkls} does not change \(P_{\ket{+}}(t)\) or \(P_{\ket{-}}(t)\), because both \(\Omega_R\hat S_x\) and \(\hat H_{\mathrm{LS}}\) are diagonal in the dressed-state basis.
The population dynamics are therefore determined only by the two dissipators and we obtain
\begin{align}
    \frac{\text{d}}{\text{d}t}\, P_{\ket{+}}(t)
    &=
    -\Gamma_\downarrow P_{\ket{+}}(t)
    +
    \Gamma_\uparrow P_{\ket{-}}(t),
    \label{eq:spinlock_meq_pplus_rate_equation}
    \\
    \frac{\text{d}}{\text{d}t}\, P_{\ket{-}}(t)
    &=
    \Gamma_\downarrow P_{\ket{+}}(t)
    -
    \Gamma_\uparrow P_{\ket{-}}(t).
    \label{eq:spinlock_meq_pminus_rate_equation}
\end{align} 
Thus \(\Gamma_\downarrow\) transfers population from \(\ket{+}\) to \(\ket{-}\), while \(\Gamma_\uparrow\) transfers population from \(\ket{-}\) to \(\ket{+}\).
Taking the difference of Eqs.~\eqref{eq:spinlock_meq_pplus_rate_equation} and \eqref{eq:spinlock_meq_pminus_rate_equation}, and using \(P_{\ket{+}}(t)+P_{\ket{-}}(t)=1\), gives
\begin{align}
    \frac{\text{d}}{\text{d}t}\,{\mathcal{S}}_c(t)
    =
    -
    \left(
    \Gamma_\downarrow+
    \Gamma_\uparrow
    \right)
    \mathcal{S}_c(t)
    +
    \left(
    \Gamma_\uparrow-
    \Gamma_\downarrow
    \right).
    \label{eq:spinlock_meq_contrast_rate_equation_general}
\end{align}
This is the most general rate equation following from the Spin-Lock master equation.
It shows that the contrast relaxes to the stationary value
\begin{align}
    \mathcal{S}_c^{\mathrm{eq}}
    =
    \frac{\Gamma_\uparrow-\Gamma_\downarrow}
    {\Gamma_\downarrow+\Gamma_\uparrow}.
    \label{eq:spinlock_meq_contrast_equilibrium}
\end{align}
For the present experiment, the nuclear bath is effectively at infinite temperature on the MHz energy scale.
We therefore have
\begin{align}
    \Gamma_\uparrow
    \simeq
    \Gamma_\downarrow,
\end{align}
such that the stationary state is the equal mixture of \(\ket{+}\) and \(\ket{-}\), and \(\mathcal{S}_c^{\mathrm{eq}}=0\).
In this limit, Eq.~\eqref{eq:spinlock_meq_contrast_rate_equation_general} reduces to
\begin{align}
    \frac{\text{d}}{\text{d}t}\,{\mathcal{S}}_c(t)
    = - \frac{1}{T_{1\rho}}
    \mathcal{S}_c(t),
    \label{eq:spinlock_meq_contrast_rate_equation}
\end{align}
where we define
\begin{align}
    \frac{1}{T_{1\rho}}
    =
    \Gamma_\downarrow+
    \Gamma_\uparrow.
    \label{eq:spinlock_meq_T1rho_definition}
\end{align}
Using Eq.~\eqref{eq:spinlock_meq_transition_rates}, this rate can be expressed as
\begin{align}
    \frac{1}{T_{1\rho}}
    =
    \frac{\gamma_{\mathrm{NV}}^2}{4}
    \left[
    S_B(+\Omega_R)
    +
    S_B(-\Omega_R)
    \right].
    \label{eq:spinlock_meq_T1rho_spectrum}
\end{align}
We have a symmetric bath spectral density, \(S_B(+\Omega_R)=S_B(-\Omega_R)\), due to the effective infinite temperature of the bath and this becomes
\begin{align}
    \frac{1}{T_{1\rho}}
    =
    \frac{\gamma_{\mathrm{NV}}^2}{2}
    S_B(\Omega_R).
    \label{eq:spinlock_meq_T1rho_symmetric_spectrum}
\end{align}

The Spin-Lock pulse starts from the dressed state \(\ket{+}\), assuming an ideal initial \(\pi/2\)-pulse.
Therefore
\begin{align}\label{eq:spin_initial_cond}
    P_{\ket{+}}(0)=1,
    \qquad
    P_{\ket{-}}(0)=0,
    \qquad
    \mathcal{S}_c(0)=1.
\end{align}
Solving Eq.~\eqref{eq:spinlock_meq_contrast_rate_equation} gives
\begin{align}
    \mathcal{S}_c(\tau)
    =
    \exp\left(-\frac{\tau}{T_{1\rho}}\right).
    \label{eq:spinlock_meq_spin_contrast_time}
\end{align}
Thus the QME predicts a simple exponential decay of the normalized spin contrast.
When the Rabi frequency is scanned, the relaxation time becomes a function of \(\Omega_R\), because the bath spectrum is sampled at the dressed-state splitting.
For a fixed Spin-Lock duration \(\tau_{\mathrm{SL}}\), the corresponding fit function is therefore
\begin{align}\label{app:eq:fitfunctionwithgeneralSB}
    \mathcal{S}_c(\Omega_R,\tau)
    &=
    \exp\biggl(-\frac{\tau}{T_{1\rho}}\biggr),\\\nonumber
    &=\exp\biggl(-\frac{1}{2}\gamma_{\rm NV}^2\tau\,S_B(\Omega_R)\biggr)\fullstop
\end{align}
%% replacement note for the appendix on imperfect spin initialization 
In this derivation we have always assumed perfect optical initialization into the $\ket{0}$ state that leads to perfect preparation of the $\ket{+}$ state from Eq.~\eqref{eq:spin_initial_cond}. 
In the experiment however, this is likely not completely achieved and we only initialize with probability $a<1$, see Ref.~\cite{PhysRevLett.106.157601}, such that the NV is prepared in \(\ket{+}\) with probability \(a\) and in \(\ket{-}\) with probability \(1-a\) at the beginning of the Spin-Locking pulse.
The measured population is then related to the ideal-initialization result \(P_{\ket{+}}^{a=1}\) by
\begin{align}
P_{\ket{+}}
=
1-a+(2a-1)P_{\ket{+}}^{a=1}.
\end{align}
However, when normalizing the data as outlined in Fig.~\ref{FigNorm}\,(D), this corresponds to the following rescaling of the spin contrast, 
\begin{align}\label{eq:rescaled_spin_contrast}
\mathcal{S}_c
=
\frac{P_{\ket{+}}-P_{\ket{-}}}{a-(1-a)}
=
P_{\ket{+}}^{a=1}-P_{\ket{-}}^{a=1},
\end{align}
provided the initialization is not fully random, i.e. \(a\neq 1/2\). Crucially, the parameter $a$ does not need to be known for this normalization, the denominator in Eq.~\eqref{eq:rescaled_spin_contrast} appears in the data before the normalization as the spin contrast far from resonance.  
We therefore conclude that due to the normalization procedure we employ, an imperfect initialization with $1/2<a\leq 1$ does not influence our results.

% ---------------------------------------------------------
% Nuclear Spin Correlation Function
% ---------------------------------------------------------
\section{Nuclear Correlation Function}
\label{app:correlationfunctions}

We now compute the magnetic noise spectrum~$S_B(\omega)$ that appears in the QME, see Eq.~\eqref{eq:spinlock_meq_final_gkls}, and consequently also in the spin contrast~$\mathcal{S}_c$ that we will ultimately fit to the Spin-Lock NMR data, see Eq.~\eqref{app:eq:fitfunctionwithgeneralSB}.
The function~$S_B(\omega)$ was defined in Eq.~\eqref{eq:spinlock_meq_two_sided_spectrum} as the Fourier transform of the magnetic noise correlation function $C_B(t)$ which we defined in Eq.~\eqref{eq:spinlock_meq_bath_correlation_definition}.
Therefore, the task at hand is to explicitly compute
\begin{align}\label{app:corr:noisespectraldensitySB(w)}
S_B(\omega)
&= \int_{-\infty}^{+\infty}\text{d}t\,e^{-i\omega t}\,C_B(t)\\\nonumber
&= \int_{-\infty}^{+\infty}\text{d}t\,e^{-i\omega t}\,{\rm Tr}_{\rm nuc} \qty{\hat B(t)\hat B(0)\,\hat{\rho}_{B}}\fullstop
\end{align}
Here, $\hat{B}(t)$ is the operator that describes the magnetic field produced by the nuclear spin ensemble at the NV center's position, and it is given in the rotating frame that was obtained by applying the unitary rotation~$\hat{U}(t)$ from Eq.~\eqref{app:rotatingframeunitaryoperator} to the lab frame Hamiltonian from Eq.~\eqref{eq:spinlock_meq_lab_hamiltonian}.
Specifically, it stems from the rotation of the lab frame Hamiltonian's term $\hbar\gamma_{\rm NV}\hat{\bm B}_{\rm N}(t)\cdot\hat{\bm S}$,
\begin{align}\label{app:eq:transformationruleforB(t)}
\hat{U}(t)&\left(\hbar\gamma_{\rm NV}\hat{\bm B}_{\rm N}\cdot\hat{\bm S}\right)\hat{U}^\dagger(t)\\\nonumber
&=\hbar\gamma_{\rm NV}\hat{U}(t)\left(
\sum_k \hat{B}_{\rm N}^k\otimes\hat{ S}_k
\right)\hat{U}^\dagger(t)\\\nonumber
&=:\hbar\gamma_{\rm NV}\hat{S}_z\otimes\hat{B}(t)
\end{align}
We now start our explicit calculation of Eq.~\eqref{app:corr:noisespectraldensitySB(w)} with the general formula for the magnetic field generated by a magnetic dipole~$\bm{m}$, as can be found in~\cite{Griffiths2017a}, yielding 
\begin{align}\label{eq:aaron:dipoledipole}
\bm{B}_{\text{dipole}}(\bm{u}) = \frac{\mu_0}{4 \pi|\bm{u}|^3} \left[ 3 \qty(\bm{m} \cdot \bm{e}) \bm{e} - \bm{m}\right] \,.    
\end{align}
Here, $\bm{u}$ is the vector from the center of the magnetic dipole to the location where the magnetic field is measured, and $\bm{e}=\bm{u}/|\bm{u}|$ is a unit vector in the direction of~$\bm{u}$.
For the magnetic dipole due to a nuclear spin such as the nuclear $^{1}$H spins in question, the magnetic moment $\bm{m}$ is given by the operator projections of the nuclear spin operator $\bm{m} = \hbar\gamma_{H}\hat{\bm I}$, such that
\begin{align}\label{eq:aaron:protondipolefield}
\hat{\bm B}_{\text{proton}}(\bm{u}) =  \frac{\hbar \gamma_\mathrm{H}\mu_0}{4 \pi|\bm{u}|^3}  \left[ 3 \qty(\hat{\bm I} \cdot \bm{e}) \bm{e} - \hat{\bm I}\right]\fullstop
\end{align}
Finally, to obtain the total magnetic field of an entire ensemble of $^{1}$H spins, we compute the linear superposition of the magnetic fields due to all surface spins, and find
\begin{align}\label{labframemagneticfieldoperatoraaron}
\hat{\bm B}_{\text{N}} = \sum_j D_j \left[ 3 \qty(\hat{\bm I}^j \cdot \bm{e}^j) \bm{e}^j - \hat{\bm I}^j\right] \comma
\end{align}
where the index $j$ labels the individual $^{1}$H spins in the bath, and where we have introduced the prefactor
\begin{align}
D_j=\frac{\hbar\gamma_{\rm H}\mu_0}{4\pi |\bm u_j|^3}
\end{align}
The individual components of~$\hat{\bm B}_{\rm N}$ are
\begin{align}
    \hat{B}_\mathrm{N}^k
    &=
    \sum_j D_j
    \left[
    3\left(\hat{\bm I}^j\cdot\bm e^j\right)e^j_k
    -
    \hat I_k^j
    \right]
    \nonumber\\
    &=
    \sum_j D_j
    \left[
    3\sum_{\alpha=x,y,z}
    e^j_{\alpha}e^j_{k}\hat I_\alpha^j
    -
    \hat I_k^j
    \right]\comma
    \label{eq:magneticfieldcomponentslabframe}
\end{align}
where the quantities \(e_k^j\), with \(k\in{x,y,z}\), denote the components of $\bm{e}^j$ in the NV coordinate system.
Next, according to Eq.~\eqref{app:eq:transformationruleforB(t)}, we need to multiply these components with $\hat{S}_k$ and subsequently rotate with operator~$\hat{U}(t)$.
The NV-part of $\hat{U}(t)$ is $\text{exp}(i\omega_0 t \hat{S}_z)$, such that under the rotating wave approximation, $\hat{S}_x$ and $\hat{S}_y$ vanish and only $\hat{S}_z$ remains.
For this reason, $\hat{B}(t)$ is simply given by the rotation of $\hat{B}_{\rm N}^z$ with the nuclear spin-part of $\hat{U}(t)$ from Eq.~\eqref{eq:spinlock_meq_rotating_hamiltonian},
\begin{align}\label{app:corr:B(t)-Operator}\nonumber
\hat B(t) &= \hat{U}(t)\hat{B}_{\rm N}^z\,\hat{U}^\dagger(t)\vphantom{\sum_n}\\\nonumber
&= \exp\biggl(+i \omega_L t \sum_j \hat{I}^{j}_{z}\biggr) \hat{B}_{\rm N}^z \exp\biggl(-i \omega_L t \sum_j \hat{I}^{j}_{z}\biggr)\\\nonumber
&= \sum_j D_j\Big[3e_x^je_z^j\left(\cos(\omega_L t)\hat I_x^j-\sin(\omega_L t)\hat I_y^j\right)\nonumber\\\nonumber
&\qquad+\,3e_y^je_z^j\left(\sin(\omega_L t)\hat I_x^j+\cos(\omega_L t)\hat I_y^j\right)\vphantom{\sum_j}\\
&\qquad+\left(3(e_z^j)^2-1\right)\hat I_z^j\Big]\fullstop
\end{align}
With this expression for $\hat{B}(t)$, we now calculate $C_B(t)$, e.g. the auto-correlation function of $\hat{B}(t)$ evaluated on the infinite-temperature thermal bath state $\hat{\rho}_{B}$,
\begin{equation}\label{app:corr:correlator}
C_B(t) = {\rm Tr}_{\rm nuc} \qty{\hat B(t)\hat B(0)\,\hat{\rho}_{B}}\fullstop
\end{equation}
For such an infinite-temperature state, the nuclear spin correlations are
\begin{equation}\label{app:corr:nuc-corr-02}
{\rm Tr}_{\rm nuc}\qty{\hat I_\alpha^j(t)\hat I_\beta^i(0)\,\hat{\rho}_{B}}=\frac{1}{4}\,\delta_{ij}\delta_{\alpha\beta}\,e^{-|t|/T_{\rm nuc}}\comma
\end{equation}
which implies no interaction between different spins in the bath.
Crucially, an exponential decay is added to Eq.~\eqref{app:corr:nuc-corr-02} to account for nuclear spin decoherence due to molecular diffusion of the $^{1}$H spins in and out of the NV sensing volume.
We call the characteristic time constant of this decay $T_{\rm nuc}$.
Note that the addition of this exponential can equally well be derived by introducing noise in the nuclear Larmor frequencies of the nuclear spins ($\omega_L\rightarrow\omega_L+\delta_\omega^j$), where the noise $\delta_\omega^j$ obeys a Lorentzian distribution centered at zero with width $1/T_{\rm nuc}$. In the continuum limit this produces exactly the exponential decay factor in Eq.~\eqref{app:corr:nuc-corr-02} upon replacing the sum over the $^{1}$H spins with a volume integral and the corresponding density.

With the correlators from Eq.~\eqref{app:corr:nuc-corr-02}, Eq.~\eqref{app:corr:correlator} becomes
\begin{align}\nonumber
C_B(t) = \frac{1}{4}\,&e^{-|t|/T_{\rm nuc}}\sum_j D_j^2\biggl(\left[3(e_z^j)^2-1\right]^2\\
&+9\cos(\omega_L t)\left[(e_x^j)^2+(e_y^j)^2\right](e_z^j)^2
\biggr)\comma
\end{align}
which can be rearranged to
\begin{equation}\label{app:eq:explicitCB}
C_B(t)=e^{-|t|/T_{\rm nuc}}\biggl(C_{\rm NMR}\cos(\omega_L t) + \frac{C_0}{2}\biggr)\comma
\end{equation}
where we introduced the zero-frequency component
\begin{equation}
C_0 = \frac{1}{2}\sum_j D_j^2\left[3(e_z^j)^2-1\right]^2
\end{equation}
and the component at the nuclear Larmor frequency $\omega_L$,
\begin{equation}\label{app:eq:coefficientCNMR}
C_{\rm NMR} = \frac{9}{4}\sum_j D_j^2\left[(e_x^j)^2+(e_y^j)^2\right](e_z^j)^2\fullstop
\end{equation}
Next, we plug Eq.~\eqref{app:eq:explicitCB} into Eq.~\eqref{app:corr:noisespectraldensitySB(w)} to compute the power spectral density $S_B(\omega)$, 
\begin{align}\label{app:eq:SBwithC0andCNMR}
S_B(\omega)&=\int_{-\infty}^{+\infty}
C_B(t)e^{i\omega t}\,\text{d}t\\\nonumber
&=\vphantom{\int}C_0L(\omega)+C_{\rm NMR}\biggl(L(\omega-\omega_L)+L(\omega+\omega_L)\biggr)\comma
\end{align}
where $L(\omega)$ is a Lorentzian function with a width given by $1/T_{\rm nuc}$,
\begin{equation}\label{eq:LorentianFunction}
L(\omega) = \frac{T_{\rm nuc}}{1+\omega^2 T_{\rm nuc}^2}\fullstop
\end{equation}
Note that if we had not added the exponential decay to the nuclear spin correlator in Eq.~\eqref{app:corr:nuc-corr-02}, then one would still obtain the same result for $S_B(\omega)$, but instead of Lorentzian functions $L(\omega)$, there would be infinitely sharp Delta functions $\pi\cdot\delta(w)$.

The master-equation result above, see Eq.~\eqref{eq:spinlock_meq_final_gkls}, and the NMR fit function, see Eq.~\eqref{app:eq:fitfunctionwithgeneralSB}, both show that the relevant spin dynamic is governed by the bath spectrum $S_B(\omega)$ evaluated at the dressed-state splitting, \(\omega=\Omega_R\).
In the HH regime relevant for Spin-Lock NMR, we have \(\Omega_R\simeq\omega_L\), and therefore the Lorentzian \(L(\omega-\omega_L)\) is evaluated close to its maximum, whereas \(L(\omega)\) and \(L(\omega+\omega_L)\) are evaluated far away from their respective maximum.
Quantitatively, at \(\omega=\Omega_R\simeq\omega_L\) and for the experimentally relevant case \(\omega_LT_{\rm nuc}\gg1\), we have
\begin{align}
\frac{L(\Omega_R)}
{L(\Omega_R-\omega_L)}
\simeq
\frac{1}{1+\omega_L^2T_{\rm nuc}^2}
\ll 1
\end{align}
and
\begin{align}
\frac{L(\Omega_R+\omega_L)}
{L(\Omega_R-\omega_L)}
\simeq
\frac{1}{1+\omega_L^2T_{\rm nuc}^2}
\ll 1\fullstop
\end{align}
Since \(C_0\) and \(C_{\rm NMR}\) are geometric dipolar coefficients of comparable order, the contributions \(2C_0L(\Omega_R)\) and \(C_{\rm NMR}L(\omega+\omega_L)\) in Eq.~\eqref{app:eq:SBwithC0andCNMR} are negligibly small compared to the resonant contribution \(C_{\rm NMR}L(\Omega_R-\omega_L)\). 
For this reason, we only keep the resonant term and drop the other two.
Together with the definition of the NMR detuning~$\Delta$,
\begin{equation}
\Delta := \omega-\omega_L\comma
\end{equation}
this results in the following expression for the noise power spectral density spectrum of the $^{1}$H spin bath:
\begin{equation}\label{app:eq:shortFormofCB(w)}
S_B(\omega) = C_{\rm NMR}\cdot L(\Delta) \fullstop
\end{equation}
What is left to do is to compute explicit expressions for the coefficient $C_{\rm NMR}$ from Eq.~\eqref{app:eq:coefficientCNMR}.
To that end, we substitute the sum over all nuclear spins with a volume integral,
\begin{equation}
\sum_j\rightarrow\rho_V\int_V\text{d}V\,
\end{equation}
where $\rho_V$ is the volume spin density of the nuclear spins in the bath.
For the evaluation of this volume integral, we use the coordinate system where $(0,0,1)$ is the surface normal vector. 
For this choice of coordinate system, the NV center's orthonormal coordinate system $\{\bm{x},\bm{y},\bm{z}\}$ can be written in the following spherical representation
\begin{align}
\bm{x} &= \left(\begin{matrix}
\sin(\pi/2)\cos(\beta-\pi/2)\\
\sin(\pi/2)\sin(\beta-\pi/2)\\
\cos(\pi/2)\end{matrix}\right)\,,\\\nonumber
\bm{y} &= \left(\begin{matrix}
\sin(\alpha-\pi/2)\cos(\beta)\\
\sin(\alpha-\pi/2)\sin(\beta)\\
\cos(\alpha-\pi/2)\end{matrix}\right)\,,\\\nonumber
\bm{z} &= \left(\begin{matrix}
\sin(\alpha)\cos(\beta)\\
\sin(\alpha)\sin(\beta)\\
\cos(\alpha)\end{matrix}\right)\comma
\end{align}
where $\bm{z}$ points along the NV axis which is tilted by an angle $\alpha$ with respect to the diamond surface normal vector $(0,0,1)$.
Furthermore, the position of a nuclear spin in the bath (over which we integrate) is
\begin{align}
\bm{u} &= \left(\begin{matrix}
r\sin(\theta)\cos(\phi)\\
r\sin(\theta)\sin(\phi)\\
r\cos(\theta)\end{matrix}\right)\,,
\end{align}
With these coordinates for the NV and the nuclear spin, the unit vector components that appear in $C_{\rm NMR}$ from Eq.~\eqref{app:eq:coefficientCNMR} are
\begin{align}
(e_x^j) &= \bm{x}\cdot\bm{u}/r\comma\\\nonumber
(e_y^j) &= \bm{y}\cdot\bm{u}/r\comma\\\nonumber
(e_z^j) &= \bm{z}\cdot\bm{u}/r\comma\\\nonumber
|\bm{u}^j|&=r\fullstop
\end{align}
For an NV at a depth~$d_{\rm NV}$ inside the diamond crystal and a nuclear spin bath that fills the entire semi-infinite volume on the diamond surface, see Fig.~\ref{Fig1}\,(A), we then obtain the following volume integral:
\begin{align}\label{app:corr:integrationCnmr}
C_{\rm NMR} &= \frac{9}{4}\int_0^{2\pi}\int_0^{\pi/2}\int_{d_{\rm NV}/\cos(\theta)}^{\infty}
\rho_V\,\frac{(\hbar\mu_0\gamma_{\rm H})^2}{16 \pi^2 r^6}\vphantom{\int^\int}\\\nonumber
&\quad\left[(\bm{x}\cdot\bm{u})^2+(\bm{y}\cdot\bm{u})^2\right](\bm{z}\cdot\bm{u})^2
r^2 \sin(\theta)\,\text{d}r\,\text{d}\theta\,\text{d}\phi\comma
\end{align}
which evaluates to
\begin{equation}
C_{\rm NMR} = \frac{\rho_V(\hbar\mu_0\gamma_{\rm H})^2}{2048\pi\cdot d_{\rm NV}^3}
\biggl(
8-3\sin(\alpha)^4
\biggr)
\end{equation}
For our diamond samples, the diamond surface is aligned along the $[100]$ crystal direction, such that
\begin{align}
\alpha=\cos^{-1}\left(\frac{(1,1,1)\cdot(1,0,0)}{|(1,1,1)|\cdot|(1,0,0)|}\right)\approx\SI{54.7}{\degree}\fullstop
\end{align}
With this, we ultimately obtain
\begin{align}\label{app:corr:CNMRfinal}
C_{\rm NMR} &= \frac{5\,\rho_V(\hbar\mu_0\gamma_{\rm H})^2}{1536\,\pi\cdot d_{\rm NV}^3}\comma
\end{align}
such that overall the magnetic noise spectrum from Eq.~\eqref{app:eq:shortFormofCB(w)} becomes
\begin{equation}
S_B(\omega)=\frac{5\,\rho_V(\hbar\mu_0\gamma_{\rm H})^2}{1536\,\pi\cdot d_{\rm NV}^3}\cdot\frac{T_{\rm nuc}}{1+\Delta^2 T_{\rm nuc}^2}\fullstop
\end{equation}
We note that for a nuclear spin bath of finite vertical thickness~$Z$, the integration limits in Eq.~\eqref{app:corr:integrationCnmr} need to be modified such that one integrates $r$ from $d_{\rm NV}/\cos(\theta)$ to $(d_{\rm NV}+Z)/\cos(\theta)$, which results in
\begin{align}\label{eq:app:cnmrforfinitelayers}
C_{\rm NMR}^{\rm layer} &= \frac{5\,\rho_V(\hbar\mu_0\gamma_{\rm H})^2}{1536\,\pi}\left(\frac{1}{d_{\rm NV}^3}-\frac{1}{(d_{\rm NV}+Z)^3}\right)\fullstop
\end{align}
Alternatively, if one considers a two-dimensional surface ensemble of surface spin density~$\rho_A$, one needs to fix $r=d_{\rm NV}/\cos(\theta)$ and integrate only over $\phi$ and $\theta$, which yields
\begin{align}
C_{\rm NMR}^{\rm surface} &= \frac{41\,\rho_A(\hbar\mu_0\gamma_{\rm H})^2}{5040\,\pi\cdot d_{\rm NV}^4}\fullstop
\end{align}
Finally, we comment that $C_{\rm NMR}$ has the dimension of magnetic field squared.
In their publication, Pham \textit{et al.} also obtain Eq.~\eqref{app:corr:CNMRfinal} and call it the root mean squared magnetic field $B_{\rm rms}^2$ of the nuclear spin fluctuations~\cite{Pham2016a}.

% ---------------------------------------------------------
% Filter Function Computation
% ---------------------------------------------------------

\section{Filter Function Computation}
\label{app:filterfunc}

Loretz \textit{et al.}~\cite{Loretz2015a} have introduced a numerical model for calculating the filter function of XY8 sequences with realistic $\pi$-pulses of finite duration~$\tau_\pi$.
Rather than deriving the filter function from the Fourier transform of the sequence modulation function as is the traditional approach in filter theory~\cite{Biercuk2011a}, their approach numerically simulates the transition probability $\ket{0}\rightarrow\ket{-1}$ of the NV spin under XY8 in the presence of nuclear spin noise.
This numerical treatment allows for the treatment of realistic non-zero pulse durations, which are difficult to incorporate within the conventional filter-function formalism.
Moreover, their model can also be applied to Spin-Lock, whose continuous microwave drive does not possess a well-defined modulation function in the usual rotating frame.
In this appendix, we use a slightly modified version of the model of Loretz \textit{et al.} to first reproduce their XY8 filter function results, and subsequently derive the corresponding Spin-Lock filter function.
The results of this are shown in the main text in Fig.~\ref{Fig2}\,(A).

We begin with considering a rotating frame Hamiltonian for the coupling of the NV spin to a nuclear spin, expressed in NV spin operators $\hat{S}_{\{x,y,z\}}$ and nuclear spin operators $\hat{I}_{\{x,y,z\}}$, where the NV is treated as an effective two-level system spanned by $\ket{0}$ and $\ket{-1}$,
\begin{align}\label{eq:filtfunc:hamiltonian}
\hat{\mathcal{H}}_\kappa(\Omega_R)=\ &
\underbrace{\vphantom{\sum}\hbar\delta\cdot\hat{S}_z}_{\text{detuning}}
+ \underbrace{\vphantom{\sum}\hbar\Omega_R\cdot\hat{S}_{\kappa}}_{\text{control sequence}}
+ \underbrace{\vphantom{\sum}\hbar\omega_{L}\cdot\hat{I}_z}_{\text{nuc. Larmor}}\\\nonumber
&+\underbrace{\vphantom{\sum}\hbar a_\perp (\hat{S}_z\otimes\hat{I}_x + \hat{S}_z\otimes\hat{I}_y)}_{\text{interaction of NV and nuc. spin}}
\end{align}
Here, the first term describes a static detuning $\delta$ of the NV spin with respect to the rotating frame of reference; the second term describes an applied driving field with Rabi frequency $\Omega_R$ along direction $\kappa\in\{x,y\}$; the third term describes the effective Larmor frequency $\omega_L$ of the nuclear spin; and the fourth term describes the transverse hyperfine interaction of the NV spin and the nuclear spin in question, where $a_\perp$ is the coupling strength. 
Hamiltonian in Eq.~\eqref{eq:filtfunc:hamiltonian} differs from what is shown in \cite{Loretz2015a} by the addition of the term $a_\perp\hat{S}_z\otimes\hat{I}_y$ to describe hyperfine interaction both transverse directions. 
%We further note that this equation is meant for the coupling of the NV spin with a single nuclear spin, not to an entire ensemble, however, the authors of \cite{Loretz2015a} claim that this is approximately valid for ensembles by setting $a_\perp=\gamma_{\rm NV} B_{\rm rms}$.

We continue with constructing the time evolution operator that describes the evolution of the system under an applied XY8 sequence. 
For this, we first build the propagator for a $\pi/2$-pulse applied in $x$ direction,
\begin{equation}\nonumber
U_{\pi/2}^{X} = e^{ -i \tau_\pi/2\ \hat{\mathcal{H}}_x(\Omega_R)/\hbar }\comma
\end{equation}
where $\tau_\pi$ is the $\pi$-pulse duration.
Next, we build spin-echo block propagators, i.e. $(\tau_{\rm gap}/2)-(\pi)-(\tau_{\rm gap}/2)$, with the $\pi$-pulse applied in either $x$ or $y$ direction
\begin{align}\nonumber
&U_{\pi}^X = e^{ -i \tau_{\rm gap}/2\ \hat{\mathcal{H}}(0)/\hbar}\cdot e^{ -i \tau_\pi\hat{\mathcal{H}}_x(\Omega_R)/\hbar}\cdot e^{ -i \tau_{\rm gap}/2\ \hat{\mathcal{H}}(0)/\hbar}\comma\\\nonumber
&U_{\pi}^Y = e^{ -i \tau_{\rm gap}/2\ \hat{\mathcal{H}}(0)/\hbar}\cdot e^{ -i \tau_\pi\hat{\mathcal{H}}_y(\Omega_R)/\hbar}\cdot e^{ -i \tau_{\rm gap}/2\ \hat{\mathcal{H}}(0)/\hbar}\comma
\end{align}
where $\tau_{\rm gap}$ is the distance between the edges of two consecutive $\pi$-pulses, such that $\tau_0 = \tau_\pi+\tau_{\rm gap}$ is the center-to-center distance between $\pi$-pulses, as depicted in Fig.~\ref{Fig1}\,(B).
The total XY8 time evolution operator is then given by concatenation of spin-echo blocks,
\begin{equation}
U_{\rm XY} = U_{\pi/2}^{X}\left[U_{\pi}^X U_{\pi}^Y U_{\pi}^X U_{\pi}^Y U_{\pi}^Y U_{\pi}^X U_{\pi}^Y U_{\pi}^X\right]^k U_{\pi/2}^{X}
\end{equation}
where $k$ is the XY8 repetition number.
The Spin-Lock time evolution operator is constructed in similar fashion, with two $\pi_x/2$-pulses and a microwave pulse along $y$ with amplitude $\Omega_R$ and duration $\tau$,
\begin{equation}
U_{\rm SL} = U_{\pi/2}^{X} \left[ e^{ -i \tau\hat{\mathcal{H}}_y(\Omega_R)/\hbar } \right] U_{\pi/2}^{X}\fullstop
\end{equation}
Next, we assume that the NV spin is initialized via optical pumping into the state $\ket{0}$, and that the nuclear spin is in a fully mixed state, such that the overall initial state at the beginning of the sequence is
\begin{equation}
\hat{\rho}_0 = 
\underbrace{\vphantom{\int}\left(\begin{matrix} 1&0\\0&0\end{matrix}\right)}_\text{NV spin}
\otimes\underbrace{\vphantom{\int}\left(\begin{matrix} 1/2&0\\0&1/2\end{matrix}\right)}_\text{nuclear spin}\fullstop
\end{equation}
The property we are ultimately interested in computing is the probability $P_{\ket{-1}}$ to find the NV spin in state $\ket{-1}$ at the end of the sequence, given that it was initially prepared in $\ket{0}$,
\begin{equation}\label{eq:filtfunc:probability}
P_{\ket{-1}} = \frac{1}{2}+\text{Tr}\left[ (\hat{S}_z\otimes\hat{I}_0)\cdot(U^\dagger\hat{\rho}_0\,U)\right]\comma
\end{equation}
where $U$ is either $U_{XY}$ or $U_{SL}$, and the addition of $\frac{1}{2}$ is to map the expectation value given by the trace (with values between $-\frac{1}{2}$ and $+\frac{1}{2}$) onto the desired probability (with values between 0 and 1).

Eq.~\eqref{eq:filtfunc:probability} is evaluated as a function of the nuclear noise frequency $f_{\rm noise}=\omega_L/(2\pi)$ and with a fixed sequence detection frequency of $f=\SI{2}{MHz}$; meaning for XY8 we fix $\tau_0=\SI{250}{ns}$ and for Spin-Lock we fix $\Omega_R=2\pi\times\SI{2}{MHz}$.
This way, we essentially simulate the NV response to different noise frequencies under an NMR sequence that is tuned to pick up noise at \SI{2}{MHz}.
For this reason, $P_{\ket{-1}}(f_{\rm noise})$ can be interpreted as an effective filter function.
Evaluation for $f=\SI{2}{MHz}$ specifically matches our experimental conditions for $^{1}$H NMR in a bias magnetic field of approximately $\SI{500}{G}$, because at this field the $^{1}$H Larmor frequency is roughly $\SI{2}{MHz}$.

The result of this calculation is shown in the main text in Fig.~\ref{Fig2}\,(A), where we use a microwave Rabi frequency in XY8 that is similar to what we work with in our experiments, $\Omega_R=2\pi\times\SI{10}{MHz}$, such that $\tau_\pi=\SI{50}{ns}$ and $\tau_{\rm gap}=\SI{200}{ns}$. 
The $\pi$-pulses therefore make up \SI{20}{\percent} of the sequence duration.
In addition, both sequences are set to have a total length of \SI{200}{\micro s}, meaning we use $\tau=\SI{200}{\micro s}$ in Spin-Lock and $k=100$ for XY8 respectively.
The detuning is set to $\delta=0$, because our experiments are conducted near the ESLAC where the microwave radiation can be applied on resonance with a single hyperpolarized NV spin transition.
Finally, we set $a_\perp=2\pi\times\SI{100}{kHz}$ for both sequences to describe nearby $^{13}$C spin in the diamond crystal~\cite{Loretz2015a}.

% ---------------------------------------------------------
% FIGURE Data Normalization
% ---------------------------------------------------------
\begin{figure*}[t]
\includegraphics{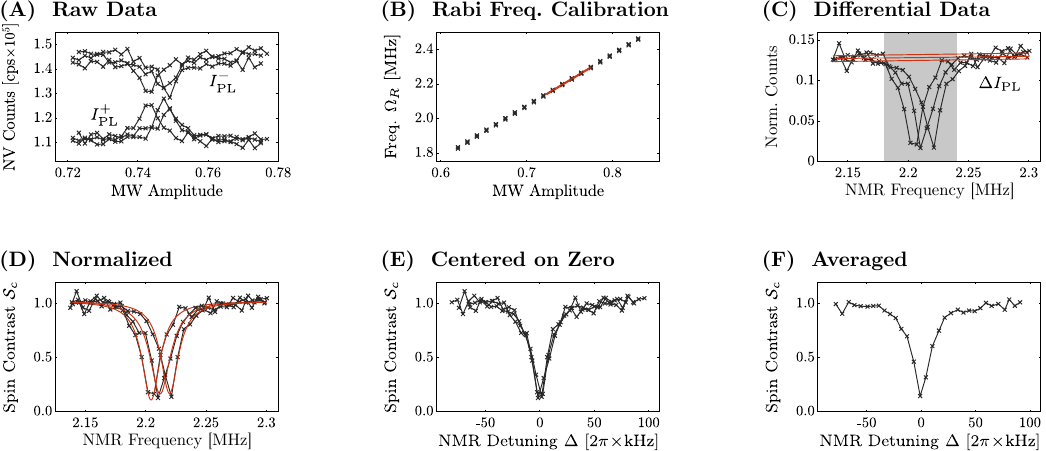}
\caption{\label{FigNorm} 
\textbf{NMR data treatment}.
\textbf{(A)}~The raw data acquired in the lab is a number of NMR spectra, given in units of NV counts and taken as a function of MW amplitude.
Here we show three such spectra as an example.
Each of them consists of two separate lines $I^{\pm}_{\rm PL}$, one for a final $+\pi/2$-pulse and one for a final $-\pi/2$-pulse.
Each spectrum is the result of exactly one hour of measurement time, and between their acquisition, we recalibrate the NV spin transition frequency and all required microwave pulses ($\pi$ and $\pi/2$) in order to eliminate the effect of slow drifts due to temperature fluctuations.
\textbf{(B)}~For each NMR spectrum, we also record a calibration of MW amplitude to Rabi frequency by measuring a series of Rabi curves and fitting the respective Rabi frequencies~$\Omega_R$.
\textbf{(C)}~As a first step, we take the difference of counts for each spectrum, $\Delta I_{\rm PL} = (I^{-}_{\rm PL}-I^{+}_{\rm PL})/(I^{+}_{\rm PL}+I^{-}_{\rm PL})$, and use each spectrum's individual $\Omega_R$ calibration to map the x-axis onto NMR frequency.
Then, we manually pick the range in which the NMR dip lies (grey area) and fit all pixels outside this range with a linear function (red).
\textbf{(D)}~Normalizing the dataset with the linear fit from (C) fit maps the y-axis onto the desired spin contrast $\mathcal{S}_c$. Next, we fit each spectrum with a Lorentzian function (red) and thereby determine the position of the NMR dip.
\textbf{(E)}~With the fitted position from (D), we plot the spectra as a function of NMR detuning, where they nicely overlap.
\textbf{(F)}~Finally, we average the spectra to obtain a single spectrum.
The same data treatment is done for XY8 NMR, with the only difference being that no calibration data is needed to translate the raw data's x-axis (in units of $\tau_0$) to NMR frequency $f=1/(2\tau_0)$.
}
\end{figure*}

% ---------------------------------------------------------
% Data Normalization
% ---------------------------------------------------------
\section{NMR Data Normalization}
\label{app:data_normalization}

In this appendix, we describe the data processing used to obtain the NMR spectra shown in Fig.~\ref{Fig2} and~\ref{Fig3}.
Specifically, we explain how the experimental microwave power control parameter is converted into the NMR detuning~$\Delta$, how the measured NV photoluminescence (PL) is transformed into the spin contrast~$\mathcal{S}_c$, and how multiple one-hour-long NMR measurements are aligned and averaged to a single multi-hour spectrum.

For each Spin-Lock NMR experiment, we acquire data using both a $(+\pi/2)$ and a $(-\pi/2)$ readout pulse, resulting in two PL traces denoted by $I_\mathrm{PL}^{\pm}$.
These traces constitute the raw experimental data, an example of which is shown in Fig.~\ref{FigNorm}\,(A).
Their x-axis is the amplitude of the applied Spin-Locking microwave field, which is controlled experimentally by the microwave signal generator through a normalized waveform amplitude between 0 and~1.

The Spin-Lock detection frequency $f=\Omega_R/(2\pi)$ is directly given by the Rabi frequency~$\Omega_R$ of the Spin-Locking pulse, e.g. its microwave amplitude.
Slow drifts in the microwave power due to temperature fluctuations and heating of microwave components (particularly the antenna) therefore cause the Spin-Lock resonance frequency to drift over time.
If left uncorrected, averaging measurements over many hours would thus artificially broaden the NMR dip.
To mitigate this effect, all measurements are acquired in one-hour segments, with the NV transition frequency~$\omega_0$ and all relevant microwave pulses (particularly the $\pi/2$ and $\pi$ pulses) recalibrated between successive runs.
Figure~\ref{FigNorm}\,(A) shows three such one-hour measurements, each containing both $I_\mathrm{PL}^{+}$ and $I_\mathrm{PL}^{-}$, corresponding to a total acquisition time of three hours.
In the following, we use this dataset to illustrate the calibration of the x- and y-axes, and the averaging of the individual spectra to a single spectrum.

As part of the recalibration between consecutive NMR measurements, we determine the relation between microwave amplitude and Rabi frequency~$\Omega_R$ by recording Rabi oscillations at several microwave amplitudes and extracting the corresponding Rabi frequencies.
The resulting calibration curves are shown in Fig.~\ref{FigNorm}\,(B) together with linear fits over the range covered by the respective NMR spectrum (red line).

In a first processing step, these linear fits are used to convert the x-axis from microwave amplitude to the NMR detection frequency, $f=\Omega_R/(2\pi)$.
At the same time, the two PL traces are combined into the relative PL difference
\begin{equation}\label{eq:nmr:deltaPL}
\Delta I_\text{PL} = \frac{ I_\text{PL}^{-}-I_\text{PL}^{+} }{ I_\text{PL}^{-}+I_\text{PL}^{+} }\comma
\end{equation}
following the definition of Pham \textit{et al.}~\cite{Pham2016a}.
The resulting spectra are shown in Fig.~\ref{FigNorm}\,(C).
By working with this relative difference $\Delta I_\text{PL}$, we suppress common-mode noise, such as slow drifts in the background PL, and ensure that $\Delta I_\text{PL}=0$ corresponds to a fully relaxed spin state.

Next, each spectrum is normalized by fitting and dividing out a linear background.
This linear background fit is performed on the data points outside a manually selected region containing the NMR dip.
This region is shown in grey in Fig.~\ref{FigNorm}\,(C), and the resulting background fits are the red lines.
The normalized spectra are plotted in Fig.~\ref{FigNorm}\,(D).
This normalization essentially removes the ``background'' relaxation due to noise sources at frequencies outside the measured NMR window (in particular zero-frequency noise), and for this reason, we identify the normalized PL values as the NV spin contrast caused by the presence of nuclear spins, denoted by $\mathcal{S}_c$ in the main text.
Accordingly, $\mathcal{S}_c=1$ indicates the absence of nuclear-spin-induced relaxation, whereas $\mathcal{S}_c<1$ reflects relaxation caused by resonant nuclear spins, and for $\mathcal{S}_c=0$ the nuclear spin noise has caused the system to relax into a fully mixed state.

Finally, the individual spectra are aligned and averaged.
Because slow microwave power drifts shift the resonance frequency between measurements, directly averaging the spectra from Fig.~\ref{FigNorm}\,(D) would artificially broaden the resulting NMR dip.
Instead, each spectrum is first fitted with a Lorentzian to determine its resonance position (red lines in Fig.~\ref{FigNorm}\,(D)), then shifted such that the dip is centered at zero detuning ($\Delta=0$), as shown in Fig.~\ref{FigNorm}\,(E), and only then averaged to obtain the final spectrum in Fig.~\ref{FigNorm}\,(F).

In summary, this procedure compensates slow drifts in the microwave power while expressing the data in the physically meaningful units $\mathcal{S}_c(\Delta)$, to which the fit function from Eq.~\eqref{eq:main_spinlock_final_fit_function} can be applied directly.

Finally, we note that the same data processing is also applied to XY8 NMR data, with one important difference.
Since the XY8 detection frequency is determined by the $\pi$-pulse spacing, $f=1/(2\tau_0)$, rather than the microwave power, no microwave calibration is required and the resonance frequency is unaffected by slow microwave power drifts.
Consequently, the segmentation into one-hour-long measurements and the subsequent spectral alignment are, in principle, unnecessary for XY8, although we nevertheless apply the same processing pipeline for consistency.

% ---------------------------------------------------------
% FIGURE SNR Optimization
% ---------------------------------------------------------
\begin{figure*}[t]
\includegraphics{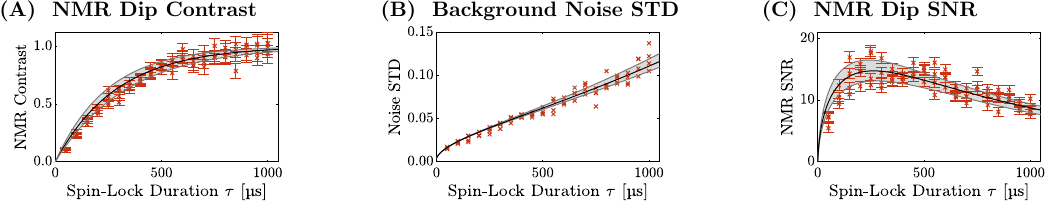}
\caption{\label{FigSNR} 
\textbf{Optimization of the SNR in Spin-Lock NMR}. We experimentally determine \textbf{(A)} the contrast $C$, \textbf{(B)} the standard deviation $\text{STD}$ of the readout noise, and \textbf{(C)} the resulting $\text{SNR}$ of a series of NMR spectra taken at different Spin-Locking durations~$\tau$.
The red crosses are experimental results, taken on NV-1, where each data point represents a single NMR experiment that ran for exactly one hour and that was normalized as described in App.~\ref{app:data_normalization}.
The error bars arise from the fitting uncertainty on the amplitude of the fitted Lorentzian lineshape.
Evaluation of Eq.~\eqref{eq:nmr:coftau}, \eqref{eq:nmr:stdoftau} and \eqref{eq:nmr:snroftau} for the experimental conditions under which the NMR data were taken results in the solid black line, and the grey area is the model's error based on the uncertainty of the employed experimental values for $T_{1\rho}^\mathrm{off}$ and $T_{1\rho}^\mathrm{res}$.
For this particular NV, SNR is maximized for $\tau\approx\SI{200}{\micro s}$.
}
\end{figure*}

% ---------------------------------------------------------
% SNR Optimization
% ---------------------------------------------------------
\section{Optimization of SNR in NMR}
\label{app:snr_optimization}

Optimizing the signal-to-noise ratio (SNR) is essential for efficient $^{1}$H NMR measurements and directly improves the sensitivity of applications such as the NV depth determination presented in this work.
Here, we discuss how to maximize the SNR by choosing an appropriate Spin-Locking duration~$\tau$.

We begin with defining the SNR as the ratio of NMR contrast $C$ and the standard deviation of the NMR spectrum's background noise, $\text{STD}(\text{background})$, yielding
\begin{equation}\label{app:snr:defsnr}
\text{SNR}:=\frac{C}{\text{STD}(\text{background})}\comma
\end{equation}
where the NMR contrast $C$ is the height of the NMR dip in units of spin contrast,
\begin{align}\label{app:snr:cdef}
C &= 1-\mathcal{S}_c(\Delta=0)\\\nonumber
&= \mathcal{S}_c(\Omega_R\neq\omega_L)-\mathcal{S}_c(\Omega_R=\omega_L)\,.
\end{align}
Importantly, we will show in the following that both $C$ and $\text{STD}$ depend on the Spin-Locking duration~$\tau$, such that there exists an optimal value of~$\tau$ that maximizes the SNR.

To identify this optimum, we consider an NV center with characteristic $T_{1\rho}^{\rm res}$ and $T_{1\rho}^{\rm off}$ times, where $T_{1\rho}^{\rm res}$ is the Spin-Lock decay time on resonance with the $^{1}$H spins ($\Omega_R=\omega_L$), and $T_{1\rho}^{\rm off}>T_{1\rho}^{\rm res}$ is the decay time off-resonance ($\Omega_R\neq\omega_L$).
In units of $|I_{\rm PL}^{+}-I_{\rm PL}^{-}|$, the corresponding spin relaxation is
\begin{align}
\Chi^\mathrm{res}(\tau) &= \Delta_{01}\,\exp\left(-\tau / T_{1\rho}^\mathrm{res}\vphantom{T_{1\rho}^\mathrm{off}} \right)\comma\\\nonumber
\Chi^\mathrm{off}(\tau) &= \Delta_{01}\,\exp\left(-\tau / T_{1\rho}^\mathrm{off}\vphantom{T_{1\rho}^\mathrm{off}} \right)\comma
\end{align}
where $\Delta_{01}$ is the PL difference between the $\ket{0}$ and $\ket{-1}$ states.
Following the normalization procedure described in App.~\ref{app:data_normalization}, the NMR contrast $C$ from Eq.~\eqref{app:snr:cdef} can be expressed as the difference of the two~$\Chi$ normalized with~$\Chi^{\rm off}$,
\begin{equation}\label{eq:nmr:coftau}
C(\tau) = \frac{\Chi^\mathrm{off}(\tau)-\Chi^\mathrm{res}(\tau)}{\Chi^\mathrm{off}(\tau)}\fullstop
\end{equation}
Next, we model the noise on the experimental readout as a Gaussian random variable with zero mean and standard deviation $\sigma$, denoted as $\mathcal{N}(0,\sigma^2)$. 
For Gaussian white noise, the standard deviation scales inversely with the square root of the number of experiment repetitions~\cite{Degen2017a}.
For a fixed total acquisition time, this corresponds to a $\sqrt{\tau}$~dependence, since fewer repetitions can be performed for longer Spin-Locking durations~$\tau$.
Furthermore, to account for the experimental overhead~$T_0$ arising from laser initialization, readout, and waiting times, we modify the scaling factor to $\sqrt{\tau+T_0}$, such that the standard deviation of the normalized NMR background yields
\begin{equation}\label{eq:nmr:stdoftau}
\text{STD}(\tau) = \text{std}\left( \frac{\mathcal{N}\left(0,[\sigma\sqrt{\tau+T_0}]^2\right)}{ \Chi^\mathrm{off}(\tau)}\right)
= \frac{\sigma\sqrt{\tau+T_0}}{\Chi^\mathrm{off}(\tau)}\fullstop
\end{equation}
Combining Eq.~\eqref{app:snr:defsnr}, \eqref{eq:nmr:coftau} and \eqref{eq:nmr:stdoftau} then gives
\begin{equation}\label{eq:nmr:snroftau}
\text{SNR}(\tau) = \frac{\Delta_{01}}{\sigma\sqrt{\tau+T_0}}\,\biggl(\exp\left(-\tau / T_{1\rho}^\mathrm{off} \right) - \exp\left(-\tau / T_{1\rho}^\mathrm{res} \right)\biggr)\,.
\end{equation}
Using Eq.~\eqref{eq:nmr:snroftau}, we now aim to find $\tau$ that maximizes the SNR, however, the equation $\text{d}/\text{d}\tau\ \text{SNR}(\tau)=0$ is transcendental, such that we cannot derive an analytic expression for this ideal $\tau$ value.
It is possible though to evaluate Eq.~\eqref{eq:nmr:snroftau} numerically for our experimental conditions, and thereby determine the value for $\tau$ that optimizes the SNR in that specific case.

For example, for NV-1, we have measured that $T_{1\rho}^\mathrm{off}=\SI{2161\pm294}{\micro s}$, $T_{1\rho}^\mathrm{res}=\SI{255\pm32}{\micro s}$, $T_0=\SI{3464}{ns}$, and $\Delta_{01} = \SI{84}{kcps}$.
Figure~\ref{FigSNR} shows the experimentally measured contrast, background noise and SNR of this NV, recorded for different values of~$\tau$, together with the predictions of Eqs.~\eqref{eq:nmr:coftau}, \eqref{eq:nmr:stdoftau}, and \eqref{eq:nmr:snroftau}.
Each data point represents one hour of data acquisition.
The grey area indicates the error on the theory prediction due to the measurement errors on the two $T_{1\rho}$ times.
The noise variance~$\sigma$ is determined by fitting Eq.~\eqref{eq:nmr:stdoftau} to the standard deviation data.

We find reasonable agreement of the experimental data with the numerical predictions.
In particular, the model predicts that for NV-1, the SNR is maximal at $\tau=\SI{262\pm22}{\micro s}$, while the data peaks at about $\tau\approx\SI{200}{\micro s}$. 
Accordingly, all Spin-Lock NMR measurements on NV-1 presented in the main text are performed using a Spin-Locking duration of \SI{200}{\micro s}.

We note that while the inclusion of the experimental overhead $T_0$ in our model ensures that the simulated STD is not vanishing for $\tau\rightarrow0$, it seems that the model is slightly overestimating the SNR for small~$\tau$.
This indicates that there are experimental circumstances that the model is currently not accounting for, such as imperfect readout~\cite{Barry2020a}, whose addition to the model could improve the overlap of the model with the SNR data.

Finally, we comment that this SNR model readily extends to XY8 NMR by replacing the relaxation times $T_{1\rho}$ with the corresponding XY8 coherence times $T_2^\mathrm{xy}$ and substituting the Spin-Locking duration~$\tau$ with the XY8 sequence duration $8k\tau_0$.
This results in a functional dependency $\text{SNR}(k)$, yielding
\begin{align}\label{eq:xy8:snrofk}
&\text{SNR}(k) = \frac{\Delta_{01}}{\sigma\sqrt{8k\tau_0+T_0}}\ \\\nonumber
&\qquad\times\left(\exp\left[\frac{-8k\tau_0}{T_{2}^\mathrm{xy,off}}\right] - \exp\left[\frac{-8k\tau_0}{T_{2}^\mathrm{xy,res\vphantom{off}}}\right]\right)\,.
\end{align}

% ---------------------------------------------------------
% Diamond Samples
% ---------------------------------------------------------
\section{Diamond Samples}
\label{app:diamondsamples}

The majority of our experimental results, that is all data except those shown in Fig.~\ref{Fig2}\,(B), were obtained on a diamond nanopillar sample provided by qnami AG (Muttenz, Switzerland).
It is an isotopically purified diamond crystal with negligible little $^{13}$C spins ($< 0.001\%$), where NVs were created by ion implantation and subsequent sample annealing.
For implantation, singly charged $^{15}$N ions with an energy of \SI{6}{keV} were used, incident at an angle of \SI{7}{\degree} relative to the surface normal, resulting in a nominal implantation depth of approximately \SI{9}{nm}~\cite{Ziegler2012a}.
To increase PL collection efficiency and for reliable identification of an individual NV defect, parabolic diamond nanopillars were fabricated~\cite{Hedrich2020a} on the diamond surface subsequent to NV creation.
Multiple pillars containing a single NV center were studied in this work.

The data shown in Fig.~\ref{Fig2}\,(B) were taken on two different diamonds.
The isotopically purified sample is nominally identical to the sample described above, but fabricated in house. 
The other diamond used in Fig.~\ref{Fig2}\,(B) is not isotopically purified, but instead contains a natural abundance of $^{13}$C spins ($\approx 1.1\%$).
The NV creation and nanopillar fabrication for this diamond were also done in house with identical recipes as described above, except for ion implantation, where we deployed $^{14}$N ions with an energy of \SI{12}{keV} corresponding to a nominal implantation depth of roughly \SI{17}{nm}~\cite{Ziegler2012a}.

% ---------------------------------------------------------
% Experiment Setup
% ---------------------------------------------------------
\section{Experimental Setup}
\label{app:experimentalsetup}
All data presented in this work were acquired on a home-built confocal microscope setup. 
An objective (Olympus LMPLFLN-100, NA = 0.8) is used to focus a green laser (Cobolt 06-MLD; emission wavelength \SI{515}{nm}) on a single NV in a nanopillar on the diamond sample surface, and to simultaneously collect the emitted red PL.
The NV is optically excited below saturation, which for our setup corresponds to a laser power of less than $\SI{70}{\micro W}$.
A static magnetic bias field is applied using a permanent neodymium disk magnet (supermagnete, GTN-32), mounted on a linear translation stage to tune the magnetic field strength at the NV location.
For precise magnetic field alignment near the ESLAC, the magnet is mounted on a goniometric stage (SmarAct SGO-60.5 and SGO-77.5).
The microwave pulses required for playing the Spin-Lock and XY8 sequences are provided by a high-frequency signal generator (Zurich Instruments SHFSG) and delivered to the NV with a gold-loop antenna.
Finally, the laser and photon detectors were gated with pulses which were created and synchronized by the same SHFSG device.

% ---------------------------------------------------------
% Bibiliography
% ---------------------------------------------------------
\bibliographystyle{libsettings.bst}
\bibliography{library.bib}

% ---------------------------------------------------------
% End of Document
% ---------------------------------------------------------
\end{document}